# Centimeter-scale synthesis of ultrathin layered MoO$_3$ by van der Waals epitaxy

*Aday J. Molina-Mendoza,[†]\* José L. Lado,[‡] Joshua O. Island,[§] Miguel Angel Niño,[∥] Lucía Aballe,[⊥] Michael Foerster,[⊥] Flavio Y. Bruno,[#] Alejandro López-Moreno,[∥] Luis Vaquero-Garzon,[∥] Herre S. J. van der Zant,[§] Gabino Rubio-Bollinger,[†∆] Nicolás Agraït,[†∆∥] Emilio M. Pérez,[∥] Joaquín Fernández-Rossier,[‡] and Andres Castellanos-Gomez[∥]\**

[†]Departamento de Física de la Materia Condensada, Universidad Autónoma de Madrid, Campus de Cantoblanco, E-28049, Madrid, Spain.

[‡]International Iberian Nanotechnology Laboratory (INL), Av. Mestre Jose Veiga, 4715-330, Braga, Portugal

[§] Kavli Institute of Nanoscience, Delft University of Technology, Lorentzweg 1, 2628 CJ Delft, The Netherlands

[∥]Instituto Madrileño de Estudios Avanzados en Nanociencia (IMDEA-nanociencia), Campus de Cantoblanco, E-28049 Madrid, Spain

[⊥]ALBA Synchrotron Light Facility, Carretera BP 1413, Km. 3.3, Cerdanyola del Vallés, Barcelona 08290, Spain

[#] Department of Quantum Matter Physics, University of Geneva, 24 Quai Ernest-Ansermet, 1211 Genève 4, Switzerland

[Δ] Condensed Matter Physics Center (IFIMAC), Universidad Autónoma de Madrid, E-28049 Madrid, Spain

ABSTRACT

We report on the large-scale synthesis of highly oriented ultrathin $MoO_3$ layers using a simple and low-cost atmospheric pressure, van der Waals epitaxy growth on muscovite mica substrates. By this method we are able to synthetize high quality centimeter-scale $MoO_3$ crystals with thicknesses ranging from 1.4 nm (two layers) up to a few nanometers. The crystals can be easily transferred to an arbitrary substrate (such as $SiO_2$) by a deterministic transfer method and be extensively characterized to demonstrate the high quality of the resulting crystal. We also study the electronic band structure of the material by density functional calculations. Interestingly, the calculations demonstrate that bulk $MoO_3$ has a rather weak electronic interlayer interaction and thus it presents a monolayer-like band structure. Finally, we demonstrate the potential of this synthesis method for optoelectronic applications by fabricating large-area field-effect devices (10 μm by 110 μm in lateral dimensions), finding responsivities of 30 mA·W$^{-1}$ for a laser power density of 13 mW·cm$^{-2}$ in the UV region of the spectrum and also as an electron acceptor in a $MoS_2$-based field-effect transistor.

INTRODUCTION

Mechanical exfoliation has paved the way to study a vast family of 2D materials isolated by cleavage of bulk layered parent crystals, such as graphene and molybdenum disulfide ($MoS_2$).[1,2] Although this technique has proven to be a very effective method to produce high quality 2D materials for fundamental research, the development of large scale synthesis methods is a crucial step towards their application in electronic and optoelectronic devices. In fact, the isolation of a new 2D material by mechanical exfoliation is usually followed by the development of new large scale synthesis techniques like chemical vapor deposition (CVD) or molecular beam epitaxy (MBE).[3,4] Molybdenum trioxide nanosheets ($MoO_3$) have been recently isolated by mechanical exfoliation of bulk α-$MoO_3$ layered crystals.[5] $MoO_3$ is a relatively low-bandgap oxide (>2.7 eV) making it very attractive for applications requiring a transparent material in the visible part of the spectrum,[6] and for devices such as field effect transistors and photodetectors.[7-9] However, a method for the large scale synthesis of this novel 2D material is still not well established.

Here we report, for the first time, the synthesis of large-scale (centimeters in lateral dimensions) and highly-oriented ultrathin (1.4 nm to few nanometers) $MoO_3$ layers, grown by an atmospheric pressure, van der Waals epitaxy growth using muscovite mica as a target substrate. We demonstrate that the as-grown $MoO_3$ crystals can be easily transferred to an arbitrary substrate (e.g. $SiO_2$, Au, Pt, quartz, polydimethylsiloxane). Through extensive characterization including X-ray photoelectron spectroscopy (XPS), low energy electron diffraction (LEED) and microscopy (LEEM), Raman spectroscopy, transmission and absorption spectroscopy, atomic force microscopy (AFM), scanning electron microscopy (SEM) and high-resolution transmission electron microscopy (HRTEM), we show that the resulting material is of high quality and suitable for device fabrication. To our knowledge, we also present the first calculations of the band structure of $MoO_3$ by density functional theory finding that the bulk material shows a similar band structure to that of the ultrathin material. We employ the grown $MoO_3$ crystals to fabricate UV-photodetectors, demonstrating that our synthesis and transfer method enables the fabrication of large-area devices. Finally, the grown $MoO_3$ crystals are also used as an electron acceptor in a mechanically exfoliated $MoS_2$-based field-effect transistor to create an open circuit voltage.

RESULTS AND DISCUSSION

$MoO_3$ is a layered semiconductor with a crystal structure that belongs to the space group *Pbnm* with the unit cell parameters listed in Table 1 (JCPDS file: 05-0508). α-$MoO_3$ consists of the stacking of double-layers (*ab* (001) and *bc* (100) projections illustrated in Figure 1a and 1b, respectively) in the *c* [001] direction with a thickness of 1.386 nm. The layers are held together by van der Waals forces to form the bulk material in a layered structure (Figure 1c), making it possible to either exfoliate and deposit or synthetize the material on a substrate by van der Waals interaction.

The $MoO_3$ crystals are grown in air using a modified *hot plate method*.[10, 11] This method, based on just placing a molybdenum foil together with a target substrate on a hot plate, is much easier and faster than other large-scale synthesis methods like the above-mentioned CVD or MBE, which require complex setups, controlled atmospheric conditions and higher temperatures than the method used here. We illustrate this process in Figure 2: a single piece of molybdenum foil (99.95% purity) is placed on a pre-heated hot plate at 540 ºC. Directly after placing the molybdenum foil on the hot plate, the target substrate used for growing the crystals is placed on top, in contact with the molybdenum foil on the hot plate, and the system is kept in this configuration for 10-60 minutes (depending on the desired thickness of the film). During this time, the molybdenum foil surface oxidizes creating a thin layer of $MoO_3$. At 540 ºC the molybdenum oxide sublimates, resulting in the deposition of $MoO_3$ on the surface of the target substrate that is at a slightly lower temperature than the molybdenum foil. Then the substrate is removed from the molybdenum foil and quenched at room temperature. This procedure can also be performed by heating a bulk $MoS_2$ flake instead of the molybdenum foil.

In the last few years there have been attempts to control the orientation of $MoO_3$ crystals during growth by changing the target growth substrate with relatively low success.[11] Here we have used mechanically exfoliated muscovite mica (referred to as mica hereafter) with an exposed (001) surface as the target substrate to fabricate highly oriented and extended $MoO_3$ layers. Figure 3 shows artistic representations of $MoO_3$ crystals grown on mica and $SiO_2$ (3a and 3b, respectively), as well as corresponding optical microscopy images of $MoO_3$ crystals grown on mica and $SiO_2$ (3c and 3d, respectively). It is important to stress that the thickness of the material is not completely homogenous and it usually ranges between 1.4 nm and 4 nm (although in some small areas it can reach 100 nm), as it will be discussed later in the AFM characterization section.

When the target substrate is $SiO_2$ (a highly-corrugated substrate with an abundance of dangling-bonds) thick and randomly oriented hexagonal $MoO_3$ crystals are grown.[5, 11] When the target substrate is mica, on the other hand, we obtain nanometer-thin α-$MoO_3$ in elongated crystals with a hexagonal shape in the *ab* plane, aligned parallel to the substrate surface and along a well-defined direction matching the symmetry of the mica surface. In fact, the cleaved (001) mica surface provides a dangling-bond-free and thus chemically inert surface that enables epitaxial growth by the so-called *van der Waals epitaxy* growth method.[12] Note that, unlike in conventional epitaxial growth, when a layered material is grown onto another layered material the lattice matching condition is drastically relaxed.[13-17] The weak van der Waals interaction with the $MoO_3$ reaching the surface during growth facilitates the diffusion of the $MoO_3$ gas on the surface, forming extended layers and hexagonal crystals of α-$MoO_3$ parallel to the mica basal plane since the in-plane interaction (chemical bonds) is stronger that of the out-of-plane interaction (van der Waals). On the other hand, when there are defects like dangling bonds or corrugation in the target surface (like in $SiO_2$), the interaction between the $MoO_3$ and the surface may be strong enough to stop the diffusion of the $MoO_3$ over the surface and thus the growth can take place in any direction and with a morphology given by unimpeded growth of the crystals faces.[18] This results in the growth of crystals with different morphologies depending on the target substrate, for example, when $MoO_3$ is grown on mica, the crystals present an elongated hexagonal shape (with two sides of the hexagon longer than the rest), while the crystals grown on $SiO_2$ present a regular hexagonal shape. By passivating the dangling bonds at the surface of other materials $MoO_3$ can be also grown onto 3D materials by van der Waals epitaxy. For example, we also observed the growth of $MoO_3$ on the $TiO_2$ terminated surface of (001) $SrTiO_3$ (STO) substrates, this results in elongated hexagonal structures oriented along the [100] and [010] axis of STO (see Figure S1 of the Supporting Information). We also demonstrate that the $MoO_3$ can be grown on exfoliated mica flakes deposited on a $SiO_2$ substrate prior the growth process (see Figure S3 of the Supporting Information)

By this process we are able to synthetize $MoO_3$ layers on the scale of tens of millimeters in lateral dimensions, finding regions of hundreds of micrometers with homogenous thicknesses (from ~1 nm to ~4nm). Thicker hexagonal crystals (tens of nanometers) with a surface of ~10 μm x 10 μm can also be found distributed on the extended thin layer. The MoO₃ extended layers can easily be grown on a larger scale (more than centimeters) by simply using a larger piece of molybdenum foil and mica substrate.

Once the growth is finished, *i.e.*, the MoO₃ crystals have covered most of the substrate, it is possible to easily separate the as-grown MoO₃ from the mica substrate using the following procedure (illustrated in Figure 4): a viscoelastic polydimethylsiloxane (PDMS) stamp (Gelfilm® from Gelpak) is placed on the substrate in contact with the MoO₃. Subsequently, the whole stack (PDMS/MoO₃/mica) is placed into a mili-Q water bath by gently peeling off the edge of the PMDS in order to permit the water is to enter between the PDMS and the mica. Due to the hydrophilic nature of mica, the water immediately penetrates between the MoO₃ and the mica substrate, completely separating the PDMS/MoO₃ from the mica. By this process, the MoO₃ remains adhered to the PDMS that floats at the air/water interface (see Figures S4 and S5 in the Supporting Information for photographs and optical microscopy images of MoO₃ on mica and on PDMS), while the mica substrate falls down (sinks) to the bottom of the water bath. It is then possible to transfer the MoO₃ from the PDMS surface to a selected substrate using the previously reported method of deterministic transfer of 2D materials.[19] In Figure 5 we show MoO₃ crystals grown on mica (Figure 5a) and then transferred to a SiO₂ substrate (Figure 5b). Because mica is a transparent material, it is very difficult to optically identify the thin material that has grown on its surface, and only the thicker crystals are visible. Once the MoO₃ is transferred to another substrate which gives higher contrast due to multiple internal reflections, such as SiO₂,[20, 21] we find that large-scale nanometer-thin MoO₃ layers are grown on the mica terraces (Figure 5c). We address the readers to Figures S6 and S7 of the Supporting Information for optical microscope images of MoO₃ transferred to other substrates. Atomic force microscope (AFM) characterization of these large-scale layers (Figure 5d) reveals a thickness ranging from 1.4 nm to 4 nm for different samples. In Figure 6 we show SEM (Figure 6a and 6b) and HRTEM (Figure 6c) images of ultrathin $MoO_3$ extended layers transferred to a TEM grid. The measured atomic distances, together with the two-dimensional fast Fourier transform (2DFFT) shown in the inset of Figure 6c, are consistent with the orthorhombic α polymorph of the $MoO_3$. The atomic distances measured by HRTEM (3.8 Å and 3.7 Å) agree with the *a* and *b* lattice parameters of $MoO_3$, indicating that the $MoO_3$ crystals grow along the ⟨001⟩ plane, parallel to the mica surface, suggesting that the crystals epitaxially grow parallel to the substrate (also in agreement with low energy electron diffraction measurements shown below). The exfoliated mica substrate provides an inert surface where the $MoO_3$ is able to diffuse thanks to the weak out-of-plane interaction with the substrate, therefore, the strong in-plane interaction of $MoO_3$ triggers the formation of crystals in lateral directions (*ab* plane). Similar growing mechanisms has been observed in other materials grown by van der Waals on the surface of freshly cleaved mica.[13, 15, 16]

XPS measurements were performed on the material grown in the same conditions but transferred to an evaporated Au substrate in order to avoid possible charging effects during the measurements. Figure 7a shows the spin-orbit split Mo 3d core level, fitted with a single doublet. The binding energies for Mo $3d_{5/2}$ and Mo $3d_{3/2}$ are 232.8 and 236.0 eV, respectively, in agreement with the reported values for molybdenum in the state $Mo^{6+}$,[22, 23] and clearly different from the binding energy and line shape of Mo 3d in $MoO_2$. These values, and the absence of other components in the Mo 3d core level, indicate the growth of molybdenum oxide in a single chemical phase identified as $MoO_3$. From an overall XPS spectrum of the sample we only find Mo, O, C (from atmospheric adsorbates) and the Au (from the substrate).

To further characterize the crystal structure of the grown material, we perform Raman spectroscopy measurements ($\lambda_{exc}$ = 633 nm, Figure 7b) on the large-scale layer shown in Figure 5c (MoO$_3$ transferred to SiO$_2$). We observe peaks in the region from 1000 to 660 cm$^{-1}$, which are attributed to Mo-O stretching peaks in the region from 470 to 280 cm$^{-1}$, Mo=O bending and peaks in the region from 250 to 100 cm$^{-1}$, and Mo-O-Mo deformation modes.[24, 25] Raman spectra allows to easily distinguish MoO$_3$ from MoO$_2$ as they present differing resonance peaks.[26] The XPS and Raman spectra, give a clear proof that the grown material is pure α-MoO$_3$.

The relative orientation of the grown MoO$_3$ crystals have been studied by directly measuring the angle between the elongated directions of different crystals in optical microscope images like the one shown in Figure 8a. Figure 8b shows a histogram of the angle formed by 68 elongated MoO$_3$ hexagons with respect to the horizontal line, shown as a pink dashed line in Figure 8a, resulting in two main angles: -10° and 50°, indicating that crystals grow in directions which form an angle of ~60° between them. These results are compared with micro low energy electron diffraction (LEED) measurements, performed in a Low Energy Electron Microscope (LEEM) after annealing at 475 K in order to desorb contaminants (LEEM images are shown in Figure S8 of the Supporting Information). After this annealing the geometry and shape of the crystallites remain unaffected as checked from the electron microscopy images. Using an aperture to perform selected area diffraction we can choose a single molybdenum oxide crystal for LEED. All MoO$_3$ crystallites studied have a LEED pattern of rectangular symmetry (Figure 8c), indicating a surface with a good degree of crystallinity, although some of them have an oblique lattice distortion of 6° with respect to a rectangular shape. We observe that some of the LEED patterns appear rotated by an angle of 58° with respect to the others (Figure 8c and Figure 8d), in good agreement with the 60° obtained from the optical images measuring the growth directions on the mica substrate. The ratio of the in-plane reciprocal lattice parameters obtained from the LEED pattern is $a/b$ =1.22 in main directions, while from the other direction (rotated 58º) the ratio is 1.16, consistent with the orthorhombic α polymorph. This experimental ratio for the surface lattice is larger than the bulk value of $a/b$=1.06. This enhanced in-plane anisotropy could be due to the interaction with the underlying mica substrate during growth, but a detailed analysis of this observation should be carried out in future works as it falls out of the scope of the present manuscript.

The optical properties of the grown material have been investigated by absorption spectroscopy measurements as well as density functional theory (DFT) calculations of the band structure. For the absorption spectroscopy measurements, the as-grown $MoO_3$ layer is transferred to a quartz substrate (selected because of its high transparency in the VIS-UV range), which has also been also measured prior to $MoO_3$ deposition (Figure 9a). The spectrum shows a sudden drop to almost cero absorption for energies lower than 2 eV (wavelengths larger than 620 nm), meaning that the material remains transparent for visible light but absorbs in the UV. The inset of Figure 9a shows the transmission spectra of the as-grown material on mica, in good agreement with the absorption spectra acquired on quartz. These results indicate that the bandgap of the material should be between 2 eV and 3 eV.

To gain understanding of our experimental findings we compare with density functional theory (DFT) calculations of both bulk and monolayer $MoO_3$. Our structural relaxations, obtained using the PBEsol functional, yield in-plane lattice constants within 2% error with respect to the bulk experimental values (see Table 1) ,[27, 28] and a negligible difference between bulk and monolayer structures. Electronic structures calculations carried out over the relaxed structures were calculated using the TBmBJ scheme, yielding similar band structures between monolayer and bulk. This is indeed confirmed by our calculations of the band structure, using the Tran-Blaha modified Becke-Johnson (TB-mBJ) functional that correctly describes the band gap of a wide variety of insulators (see Figure 9b for bulk $MoO_3$ and Figure 9c for the monolayer).[29] Both for bulk and monolayer, we obtain an indirect (M-Gamma) bandgap of 2.2 eV and a higher energy direct gap at Gamma point of 3.3 eV. Therefore unlike most studied 2D semiconductors, where the reduction of the thickness is usually followed by a sizeable increase of the bandgap due to quantum confinement in the out-of-plane direction, bulk $MoO_3$ has a monolayer-like band structure. We address the reader to Figure S9 of the Supporting Information for the calculated optical conductivity and density of states of $MoO_3$.

In order to explore the potential applications of the grown $MoO_3$ crystals for large-area devices we fabricate and study field-effect devices based on extended $MoO_3$ layers transferred onto a 285 nm thick $SiO_2$ substrate. The $SiO_2$ is thermally grown on a highly doped Si(p+) substrate which is used as back-gate. 110 x 110 $\mu m^2$ Ti/Au electrodes are evaporated on the $MoO_3$ using a shadow mask in order to fabricate the devices (Figure S10 in the Supporting Information). The shadow mask results in cleaner devices which have not been exposed to lithography resists. The electrodes are separated by 10 $\mu m$ from each other, yielding large area (110 x 10 $\mu m^2$) $MoO_3$-based devices. Note that the area of our $MoO_3$-based devices is up to 2000 times larger than the area of previously reported $MoO_3$ devices.[7, 9] Figure 10a and Figure 10b show both optical microscope and AFM topographic images of one of these devices in which the area between the two electrodes is completely covered by a very thin layer of $MoO_3$ (~1.4 nm - 6 nm). The field-effect characteristics of the device are determined in combination with photocurrent generation of the device upon illumination with a 405 nm wavelength laser. Figure 10c shows the transfer curves of the device shown in Figure 10a and Figure 10b in dark conditions and under illumination at an applied source-drain voltage ($V_{bias}$) of 500 mV. The inset shows the current-voltage characteristics at an applied back-gate voltage of +40V (estimated resistivity of ~165·10$^{-3}$ $\Omega$·m). There is a weak dependence of the drain current with the back-gate voltage (small ON/OFF ratio), which indicates a high *n*-doping (MoO$_3$ is well-known to be a good electron acceptor used for surface *p*-doping).[30, 31] This is also in agreement with the current *vs.* back-gate voltage trace that clearly shows an *n*-type conduction as the current increases with increasing back-gate voltage. We calculate the photocurrent ($I_{ph}$) as the difference between the current measured in dark conditions and upon illumination. From the photocurrent we can calculate the responsivity, plotted in Figure 10d, of the device as $R = I_{ph} / P_{eff}$, where $P_{eff}$ = 14 µW is the effective power of the laser that arrives at the device ($P_{eff} = P_{laser} \cdot A_{device}/A_{spot}$). The responsivity is a typical figure-of-merit for photodetectors and represents the input-output gain of the device: the amount of photocurrent produced given the effective power of the laser. The maximum measured responsivity value is 30 mA·W$^{-1}$ at an applied source-drain voltage of 500 mV, in good agreement with the values reported for MoO$_3$ small area devices.[9, 32] Although other 2D materials have demonstrated higher responsivities in the UV (19.2 A·W$^{-1}$ with mechanically exfoliated GaS or 890 A·W$^{-1}$ for HfS$_2$)[33, 34] our devices are the first large-scale ultrathin UV-photodetectors reported up to date. In Figure S11 of the Supporting Information we can see the time response of a large-scale MoO$_3$-based photodetector, where we find a rise time of ~ 20 s and a fall time of ~ 68 s. These results make our fabricated MoO$_3$-based devices an interesting candidate for applications like visible light-transparent coatings that are sensitive to UV radiation or smart windows with absorption in the UV.

More interesting applications of MoO$_3$ are found as an electron acceptor for *p*-doping different materials.[35-37] The large work function of MoO$_3$ allows the material to drive spontaneous electron transfer from different two-dimensional materials to the MoO$_3$ crystals.[30, 38] Here we employ the as-grown MoO$_3$ crystals to modify the performance of a MoS$_2$ FET by transferring a MoO$_3$ crystal on the source electrode of the MoS$_2$ FET (van der Waals heterostructure). The result is a depletion of the electron charge carriers in the device that causes an increase in the work function.[39] The charge transfer in one of the electrodes in an MoS$_2$ FET can result in the creation of different Schottky barriers for each electrode.[40] When the FET is illuminated in order to study the photoresponse of the device, the electrons in the electrode with higher hole density are photo-excited over the barrier, generating a net current even at zero bias (like a diode). The accumulation of electrons and holes in each electrode also causes an open circuit voltage ($V_{oc}$).[40] In our device, the MoO$_3$ is transferred covering part of the MoS$_2$ flake (Figure 11a) causing an electron transfer from MoS$_2$ to MoO$_3$. When the device is illuminated with a light beam with 640 nm wavelength (photon energy above the MoS$_2$ bandgap), we see an open circuit voltage in the current-voltage characteristics of $V_{oc} \sim 8$ mV (Figure 11b). We also see this effect when investigating the time response of the device upon illumination with a light modulated intensity (Figure 11c) at zero drain-source voltage. As can be seen, there is current at zero voltage when the light is switched on and there is not current when the light is switched off. These results support the use of our grown MoO$_3$ crystals as an easy and effective method for causing electron transfer and thus the modification of Schottky barriers in functional devices which can be exploited to fabricate self-driven (working with zero bias voltage) photodetectors.

CONCLUSIONS

In summary, we have developed a technique to easily synthetize, via van der Waals epitaxy, large-scale (centimeters in lateral dimensions) and ultrathin (from 1.4 nm to few nanometers) MoO$_3$ crystals. The synthesis is carried out by sublimation of the material on a hot plate at relatively low temperature (540 ºC) and ambient pressure (without the need of a controlled atmosphere). The crystals can be easily transferred to an arbitrary substrate (SiO$_2$, Au, Pt, quartz…) using a deterministic transfer method. Different techniques (XPS, LEED and LEEM, Raman spectroscopy, transmission and absorption spectroscopy, AFM, SEM and HRTEM) have been used to characterize the quality of the as-grown material, yielding results consistent with the orthorhombic α polymorph (α-MoO$_3$). We provide the first calculations of the band structure of MoO$_3$ by density functional theory finding that the bulk material shows a monolayer-like band structure. The extended layers of ultrathin crystals have been used to fabricate field-effect devices and to study the optical response of the material by illumination with a 405 nm wavelength laser, finding responsivity values as high as 30 mA·W$^{-1}$ for a power density of 13 mW·cm$^{-2}$. These results open the door for interesting applications of large-scale and ultrathin transition metal oxides photodetectors in fields like transparent coatings or smart windows with absorption in the UV. Finally, the grown MoO$_3$ crystals are also used as an electron acceptor in a mechanically exfoliated MoS$_2$-based field-effect transistor to create voltage self-driven photodetectors.

MATERIALS AND METHODS

**$MoO_3$ crystals growth:** a molybdenum foil (99.95% purity) is used as material source and muscovite mica as target substrate. The growth is carried out on a *thermo scientific* hot plate in air (room temperature and ambient conditions) by a modified *hot plate method*.[10, 11] The growth temperature is set to 540 ˚C before starting the process and kept at that temperature during the whole procedure. The time can vary depending of the quantity of material, but the results shown in the main text are obtained after 20 minutes of growth. The target substrate, *i.e.*, the mica is then removed from the hot plate and quenched during 2-3 minutes on a glass slide.

**$MoO_3$ crystals transfer to other substrates:** a polydimethylsiloxane (PDMS) stamp (Gelfilm® from Gelpak) is used to hold the grown material once it is separated from the target substrate (mica). Mili-Q water in a pyrex beaker at room temperature is used as bath to separate the grown material from the target substrate. Once the PDMS stamp is taken out of the bath, it is blow dried with $N_2$ gas. The material is then transferred to a selected substrate using a deterministic transfer setup.[19]

**Characterization (Optical. XPS. Raman. LEED. AFM. Absorption. SEM. TEM):** optical characterization of the samples has been done with a Nikon Eclipse Ci optical microscope in reflection mode. For the X-ray Photoelectron Spectroscopy (XPS) measurements we used the Al Kα line (hv = 1486.7 eV) from a conventional monochromatized X-ray source, a hemispherical energy analyzer (SPHERA-U7) with the pass energy set to 20 eV to have a resolution of 0.5 eV. The core level spectra were fitted to mixed Gaussian−Lorentzian components; all the binding energies are calibrated using the Au 4f core level of the substrates as reference. Raman spectra were acquired on a Bruker Senterra confocal Raman microscopy instrument, using 633 nm laser excitation. Micro diffraction measurements were done at the LEEM microscope (Elmitec, GmbH) in operation at ALBA synchrotron (Barcelona, Spain).[41] The instrument combines real space imaging (LEEM) and diffraction (LEED). Using micro-LEED mode we restricted to circular regions of 5 μm diameter, allowing to choose a homogeneous area (same thickness or chemical composition) of the molybdenum oxide crystallites. Absorption spectra were acquired on an Agilent-Cary 5000 UV-VIS-NIR spectrophotometer. Atomic force microscopy (AFM) characterization was done using a (Digital Instruments D3100 AFM) operated in the amplitude modulation mode. Transmission electron microscopy (TEM) images were obtained with JEOL-JEM 3000F (1.7 Å resolution) instrument operating at 300 kV. Scanning electron microscopy (SEM) images were obtained with Zeiss EVO HD15 instrument operating at 25 kV.

**Density functional theory calculations:** DFT calculations were performed using Quantum Espresso[42] and Elk.[43] Structures for bulk and monolayer $MoO_3$ were fully relaxed with Quantum Espresso with PAW pseudopotentials and PBEsol[44] exchange correlation functional. With the relaxed structures, all electron calculations were performed with Elk, using the TB-mBJ potential,[29] specially suitable for accurate bandgap calculations. The optical spectra was calculated with RPA without local contributions in the q→0 limit as implemented in Elk, using a 20x20x1 kmesh for monolayer and 10x10x6 for bulk.

**Field-effect and optoelectronic characterization:** field-effect characteristics and optoelectronic characterization is performed in a *Lakeshore Cryogenics* probe station at room temperature and high vacuum ($<10^5$ mbar). The photoresponse is measured with a diode pump solid state laser of 405 nm wavelength operated in continuous mode and guided with an optical fiber which yields a spot of 200 μm diameter.

FIGURES

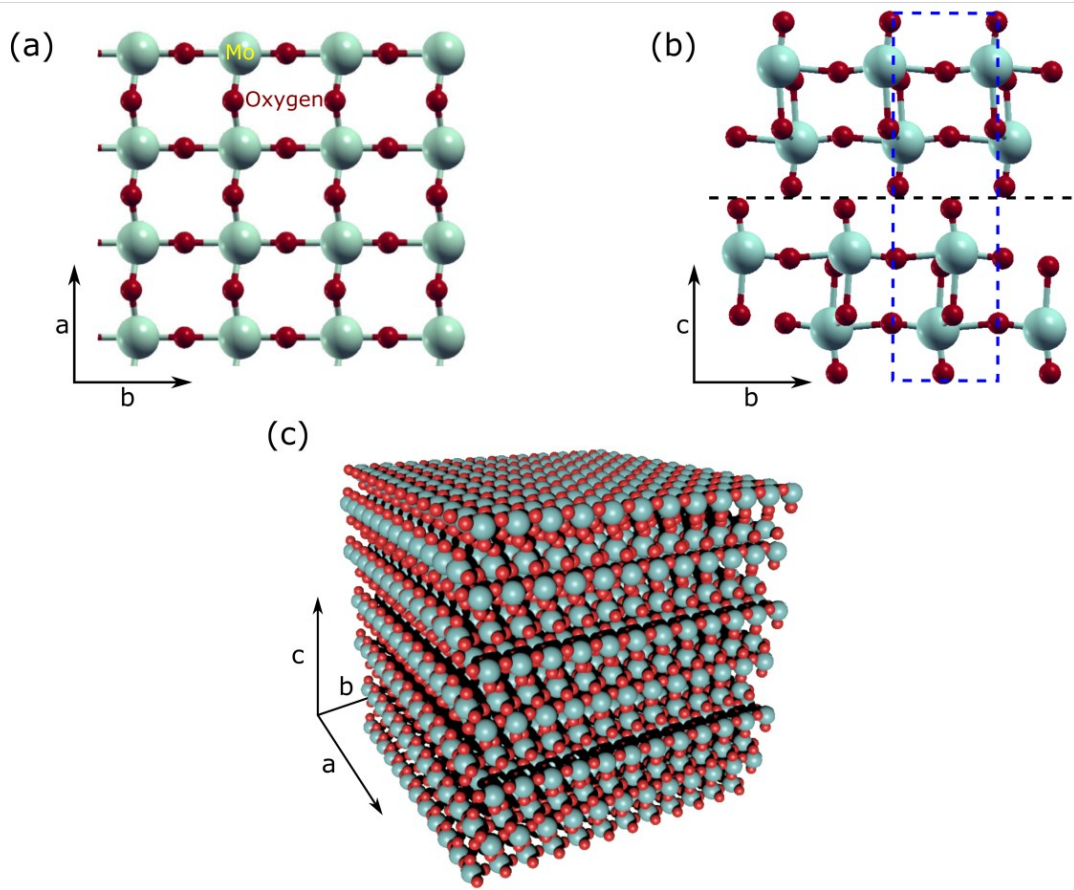

**Figure 1 (a)** *ab* plane projection of the MoO$_3$ monolayer crystal structure. **(b)** *bc* projection of the MoO$_3$ crystal structure. MoO$_3$ layers stack in the *c* direction interacting by van der Waals forces. The dashed black line indicates the *ab* plane where the interaction is via van der Waals forces and the dashed blue line indicates the unit cell. **(c)** Bulk stacking of the MoO$_3$ layers shown in (a) and (b).

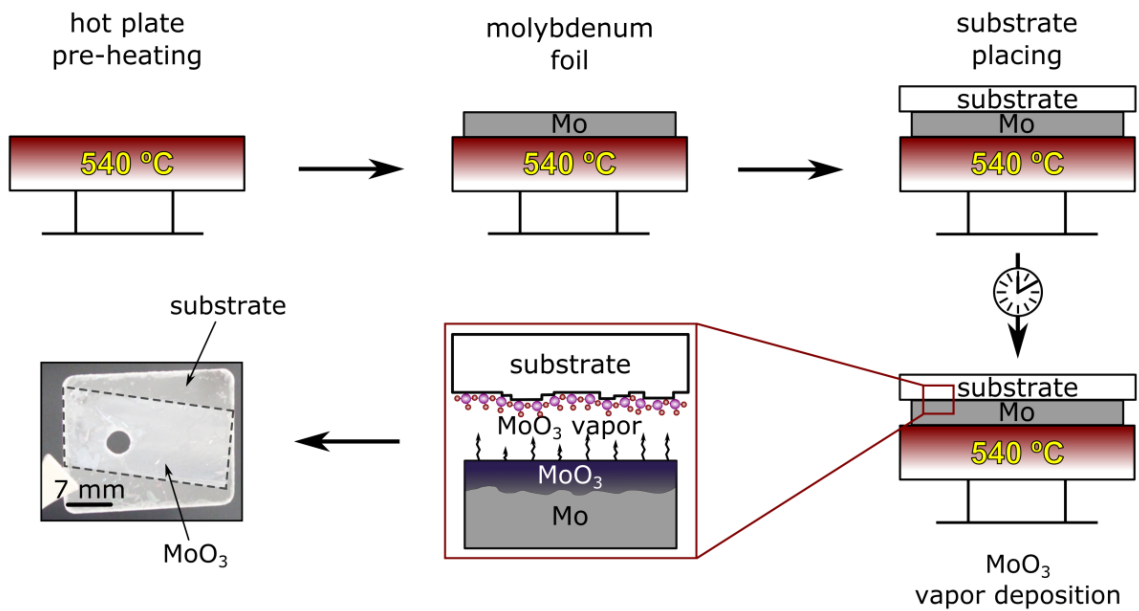

**Figure 2** Schematic drawing of the method used for growing MoO₃ in air. A piece of molybdenum foil is placed on a pre-heated hot plate at 540 ºC with a receiving substrate on top. The system is kept in this configuration for some time while the molybdenum surface oxidizes and deposits material on the substrate. The substrate is then removed and quenched at room temperature.

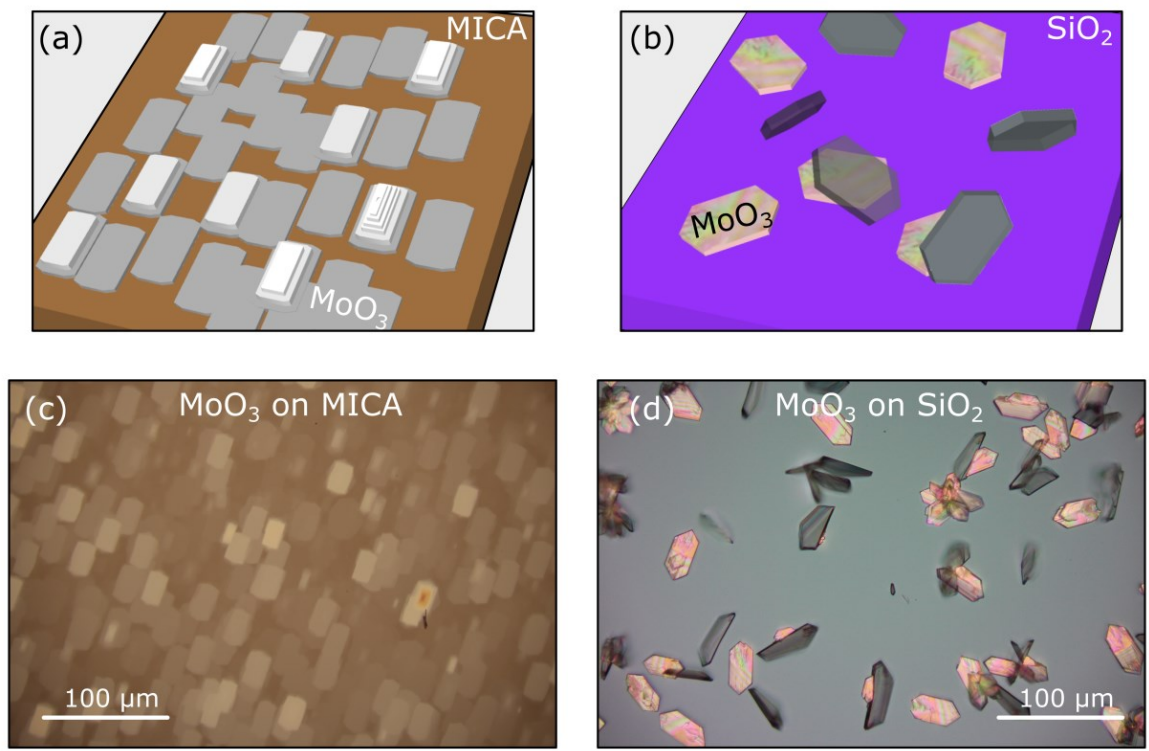

**Figure 3** Artistic representation of MoO$_3$ crystals grown on **(a)** mica and **(b)** SiO$_2$. Optical microscope images of MoO$_3$ crystals grown on **(c)** mica and **(d)** SiO$_2$. MoO$_3$ crystal dimensions (area and thickness) and ordering depend on the substrate used for growing the material via the *modified hot plate method* (Figure 2).

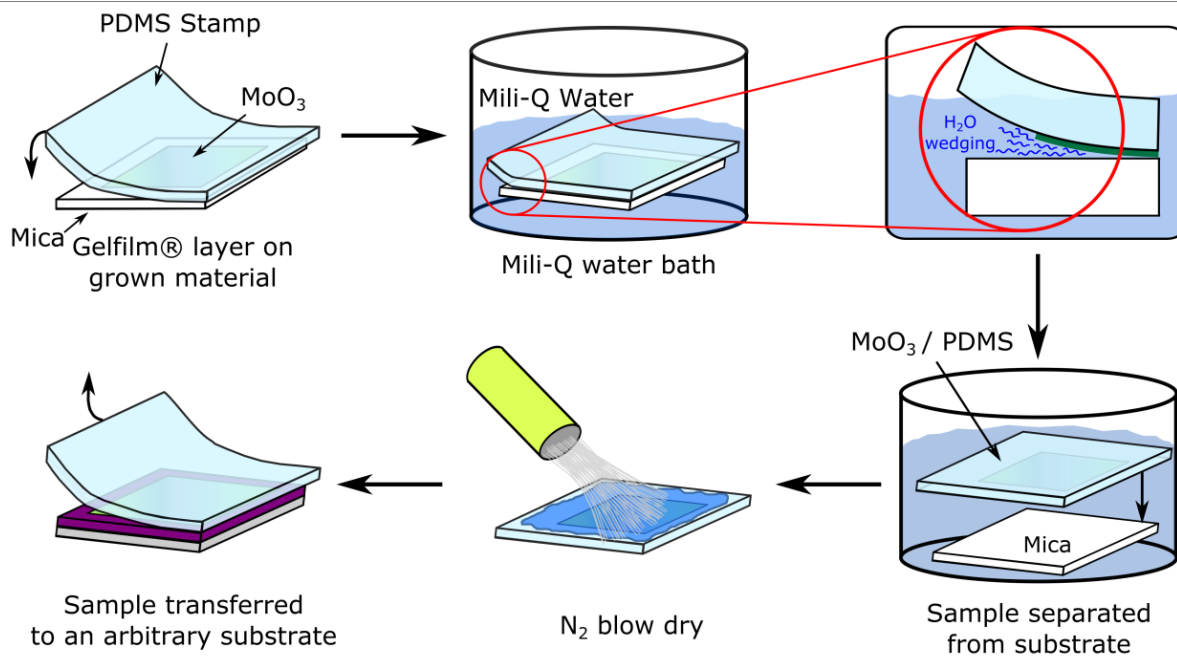

**Figure 4** Schematic drawing of the method used to transfer the as-growth MoO$_3$ crystals from mica to another substrate. A PDMS stamp (Gelfilm® from gelpak) is placed on the MoO$_3$/mica sample, sandwiching the MoO$_3$ material between the mica substrate and the PDMS stamp. The whole set (PDMS/MoO$_3$/mica) is then introduced in a Mili-Q water bath and the stamp is slightly peeled off at one of the corners. As the water enters in between the PDMS stamp and the mica substrate, the mica immediately separates from the PDMS stamp and the MoO$_3$ deposited material due to its hydrophilic behaviour, leaving all the MoO$_3$ material attached to the PDMS stamp. After blow drying with N$_2$, the MoO$_3$/PDMS is ready to be transferred to the desired substrate.

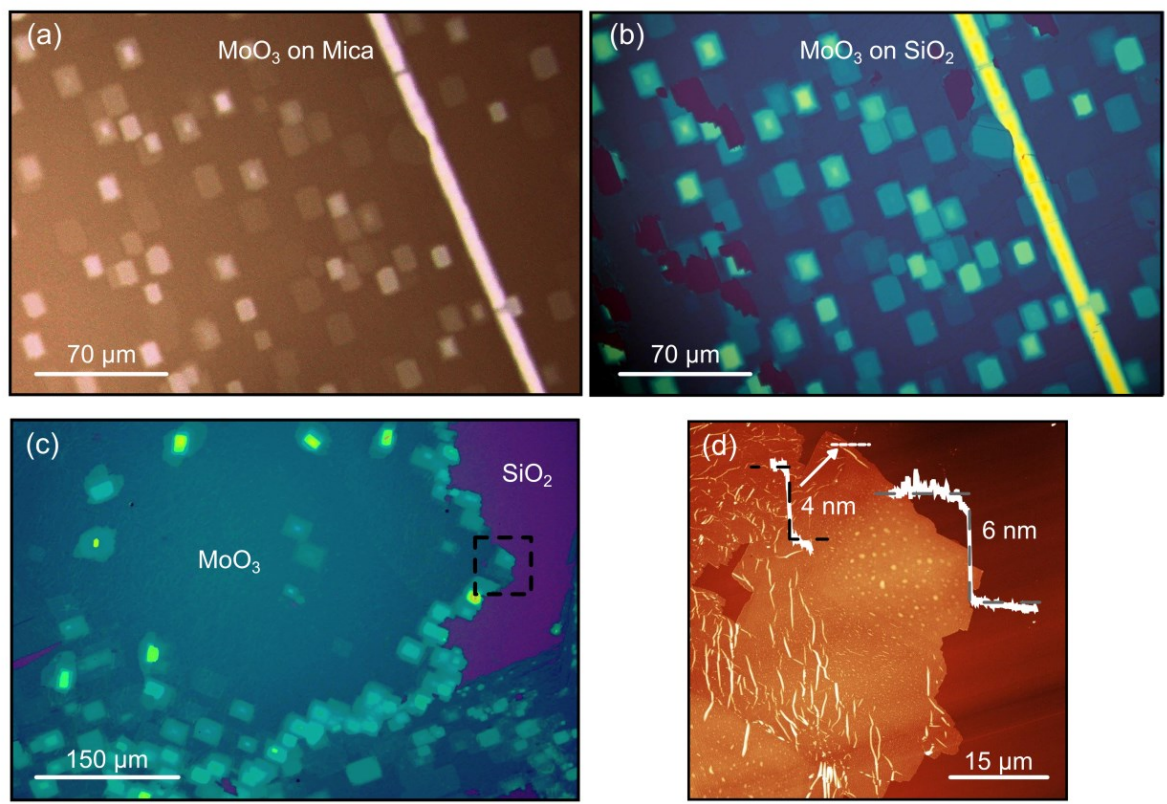

**Figure 5 (a)** Optical microscope image of MoO$_3$ crystals grown on a mica substrate. **(b)** Optical microscope image of the same MoO$_3$ crystals shown in (a) after been transferred to a SiO$_2$ substrate. **(c)** Optical microscope image of an extended MoO$_3$ single-crystal transferred to a SiO$_2$ substrate. **(d)** AFM topographic image of the area highlighted by a black square in (c). The extended layer thickness is 4 nm (~ 5 MoO$_3$ layers), while the tetragonal crystals thickness is 6 nm (~ 8 MoO$_3$ layers).

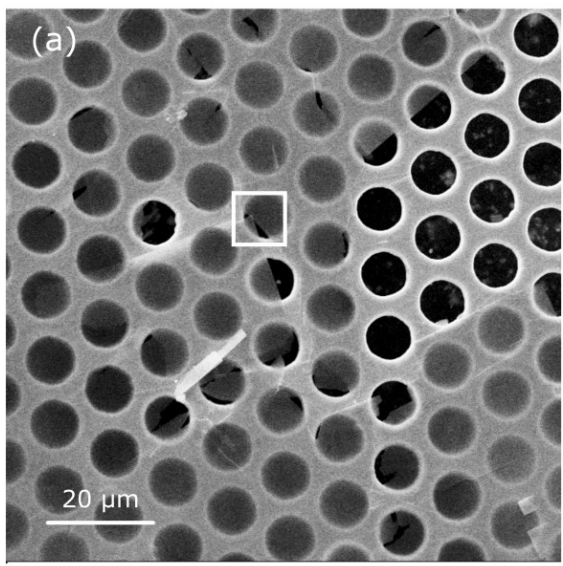
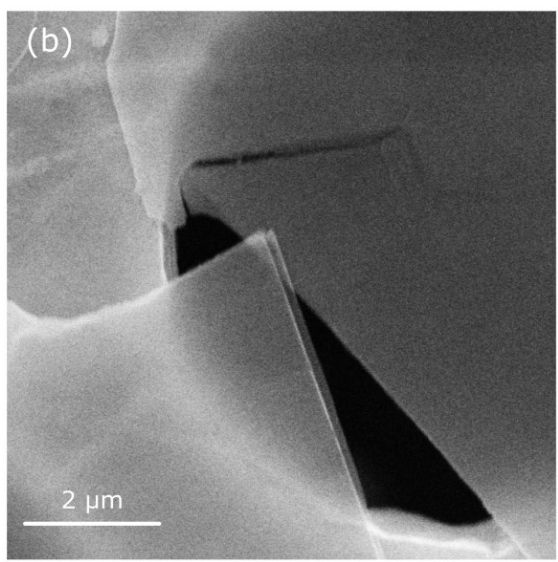
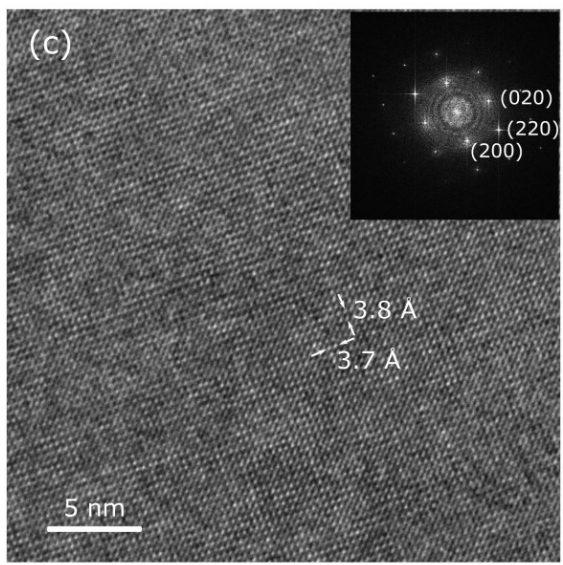

**Figure 6 (a)** SEM micrograph of several layered ultrathin MoO3 crystals deposited on a TEM grid. **(b)** Zoom in on the area marked with a white square in (a) showing a ultrathin MoO3 crystal. **(c)** Representative HRTEM of the MoO3 ultrathin layers. The atomic distances and 2DFFT (inset) are consistent with an orthorhombic α polymorph.

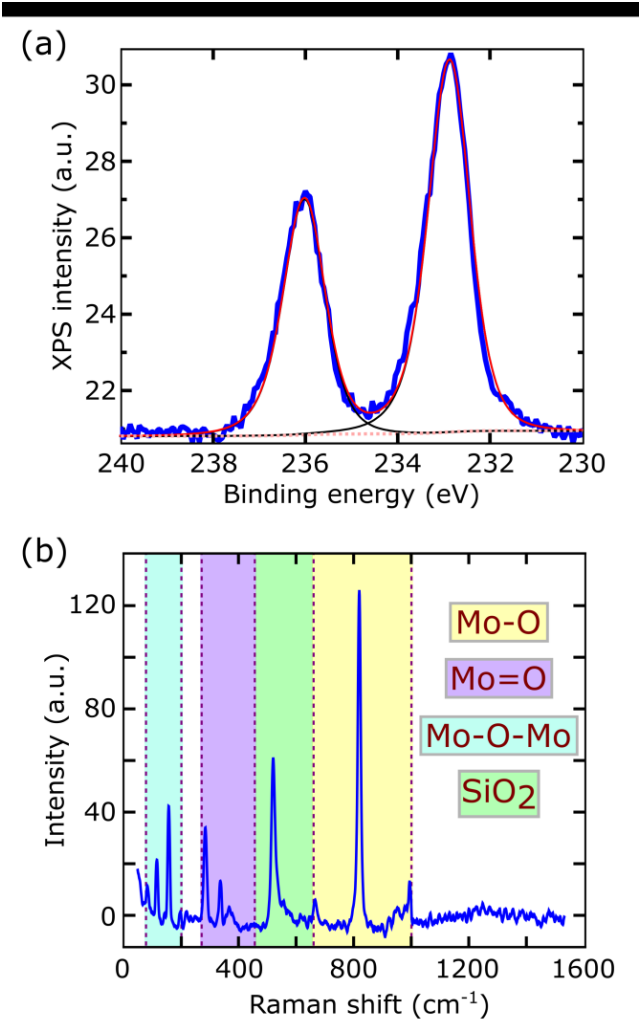

**Figure 7 (a)** XPS spectrum of Mo 3d. The blue curve represents the experimental data, while the black lines represent the fitted curve for each peak. The sum of the fitted curves for the two curves is represented as a red line, while the background is represented by the pink dashed line.

**(b)** Raman spectra acquired in the extended layer of Figure 5c. The different frequency regions are attributed to Mo-O stretching (1000-660 cm$^{-1}$), Mo=O bending (470-280 cm$^{-1}$) and Mo-O-Mo deformation modes (250-100 cm$^{-1}$).[25]

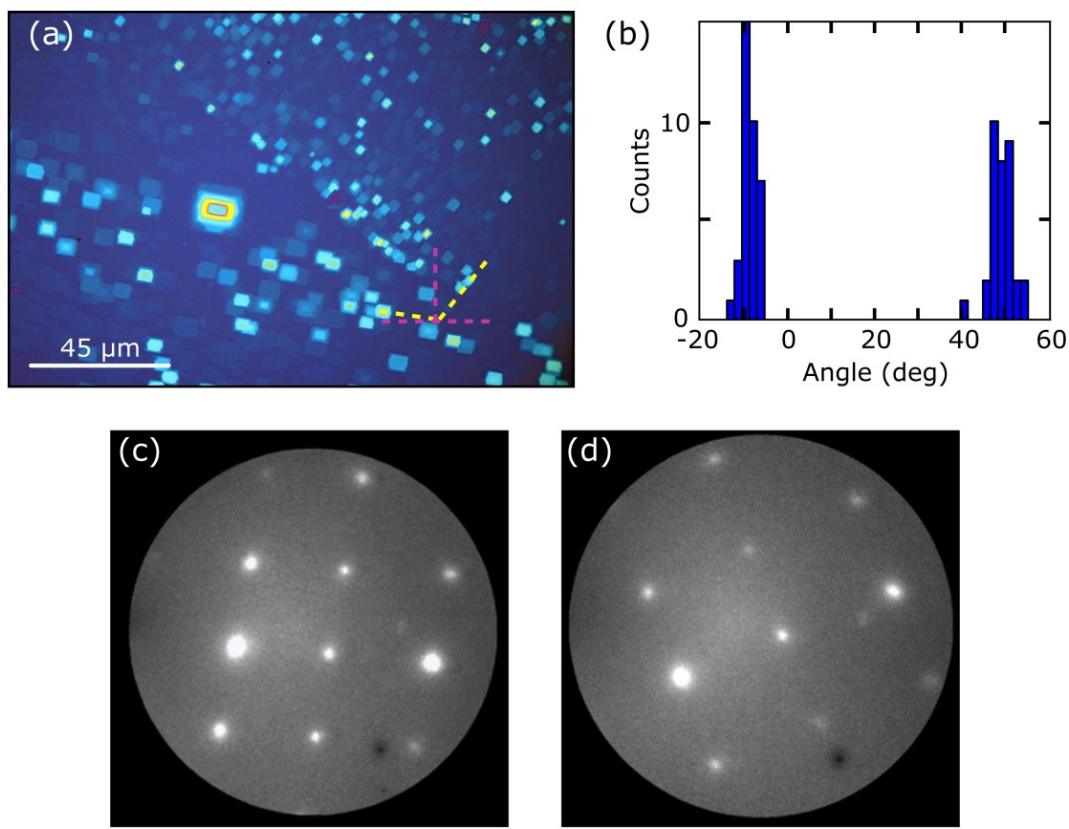

**Figure 8 (a)** Optical microscope image of MoO$_3$ crystals grown on mica and transferred to SiO$_2$. The dashed yellow lines represent the two main directions followed by the crystals during growth. **(b)** Histogram of the angle between the elongated directions followed by MoO$_3$ crystals and the horizontal directions (pink dashed lines in (a)). The two main angles are 50° and -10°, i.e., forming an angle of 60° between the two directions. **(c)** and **(d)** Micro-spot LEED pattern (E=50 eV) of different MoO3 crystals on the surface.

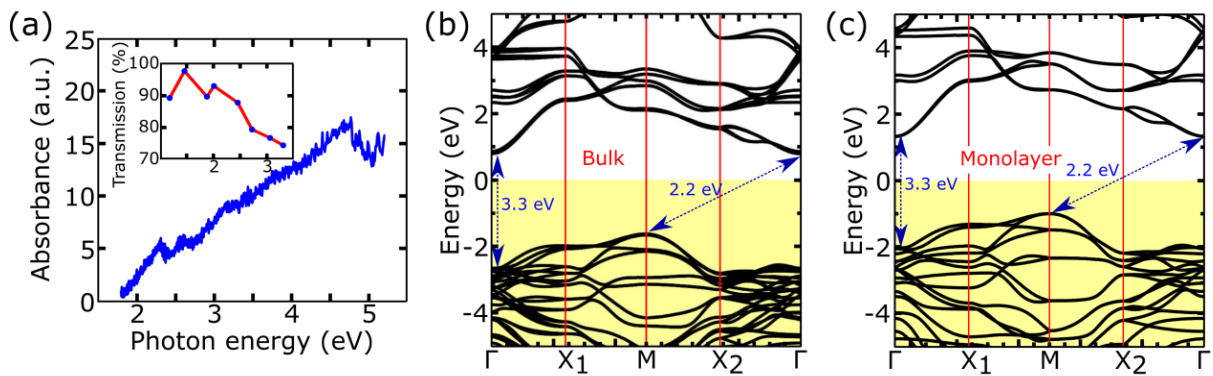

**Figure 9 (a)** Absorption spectra of MoO3 crystals transferred to a quartz substrate. Inset: transmission spectra of as-grown MoO3 crystals on mica. For energies lower than 2 eV the absorption is almost negligible. **(b)** Calculated band structure for bulk MoO3 and **(c)** monolayer crystal. The calculated optical bandgaps are 2.2 eV (indirect) and 3.3 eV (direct). The special points are defined as $X_1 = b_1/2$, $X_2 = b_2/2$ and $M = (b_1+b_2)/2$, where $b_1$ and $b_2$ are the reciprocal lattice vectors of bulk (monolayer) MoO3.

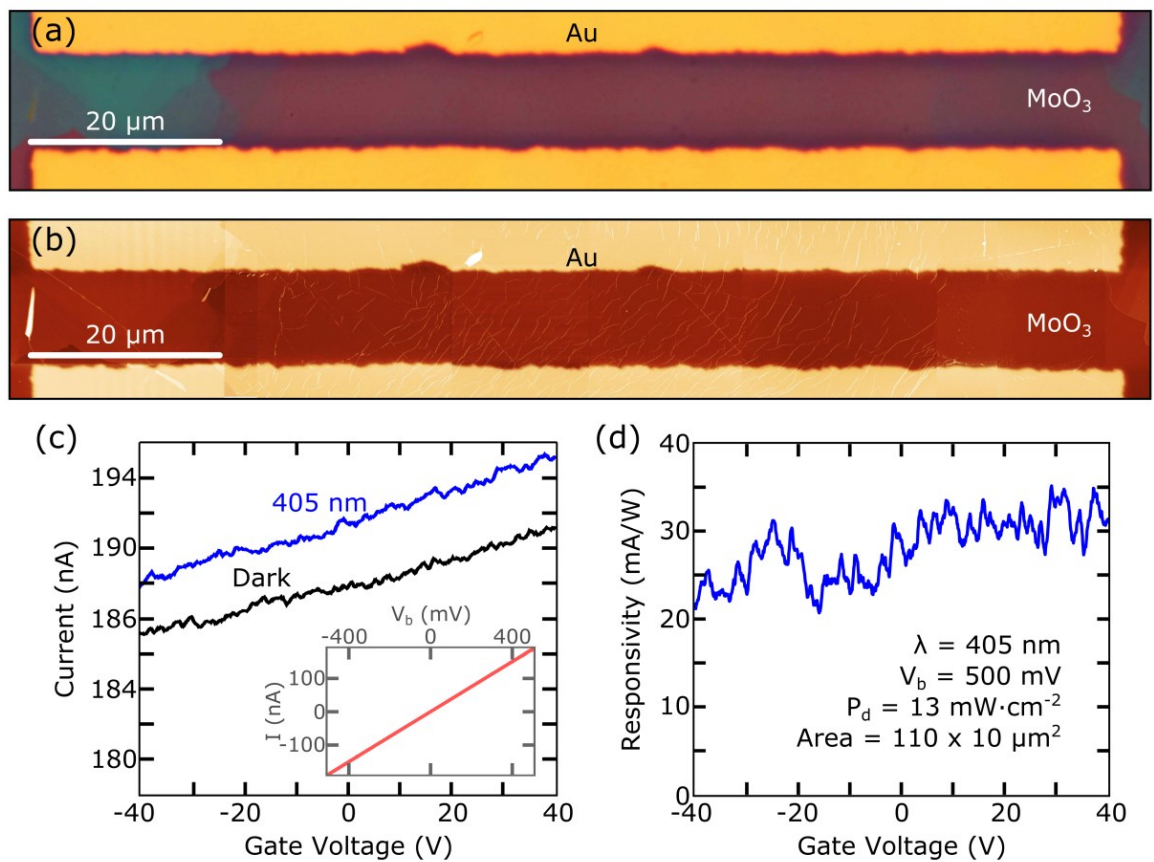

**Figure 10** MoO$_3$-based field-effect device fabricated by evaporating Ti/Au electrodes via a shadow mask on an extended layer of MoO$_3$ grown on mica and transferred to a SiO$_2$/Si(p+) substrate. **(a)** Optical microscope and **(b)** AFM topographic images of one of the devices. **(c)** Current-back gate voltage curve of the device shown in (a) measured in the dark (black) and under 405 nm wavelength light illumination (blue). Inset: Current-bias voltage curve of the same device in the dark and with an applied back-gate voltage of +40V. **(d)** Responsivity calculated for the blue curve in (a) as a function of the applied back-gate voltage.

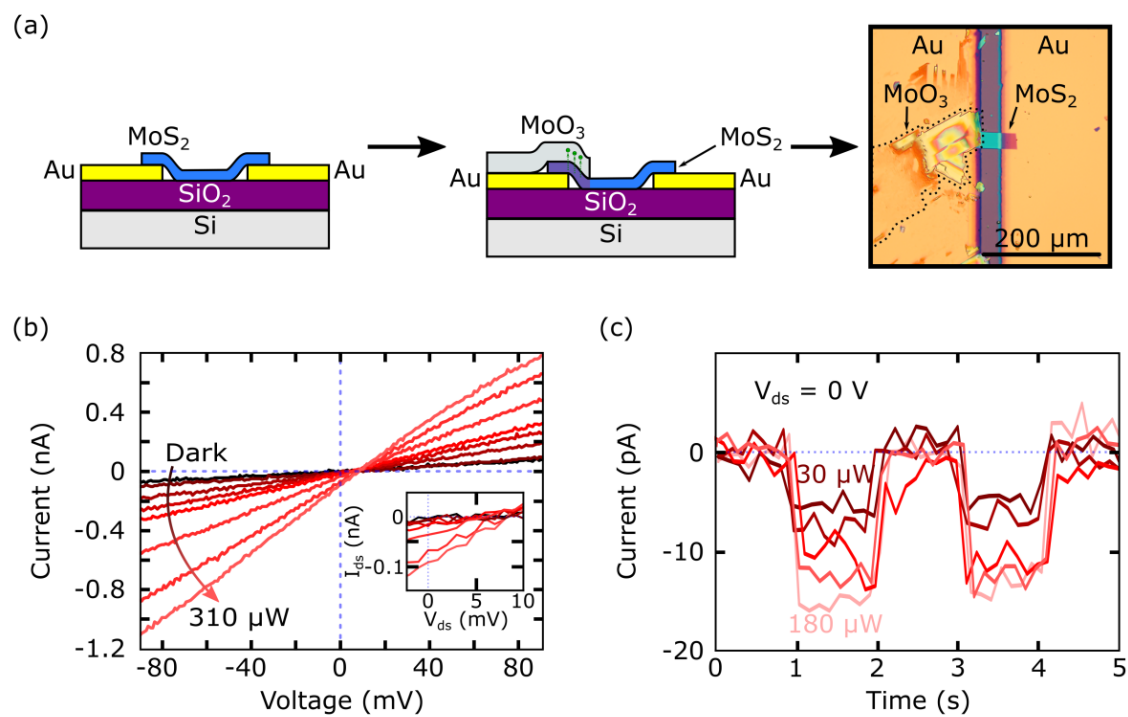

**Figure 11 (a)** Artistic representation of the $MoO_3$-$MoS_2$ heterostructure. From left to right: a $MoS_2$ field-effect transistor is fabricated by transferring a $MoS_2$ flake on pre-patterned Au electrodes on a $SiO_2$/Si substrate. Then, a $MoO_3$ crystal is transferred on one of the electrodes, partially covering the $MoS_2$ flake, causing an electron transfer from the $MoS_2$ to the $MoO_3$. On the right we show an optical microscopy image of the fabricated van der Waals heterostructure. **(b)** Current-voltage characteristics in dark conditions and upon illumination of the device for increasing light power with light wavelength of 660 nm. Inset: zoom of the area around zero drain-source voltage where the $V_{oc}$ and the $I_{sc}$ can be depicted. **(c)** Time response of the device at zero drain-source voltage in dark conditions (zero current) and upon illumination with different modulated light intensity.

TABLES

**Table 1.** MoO$_3$ lattice parameters.

| Material dimension | a (Å) | b (Å) | c (Å) |
|---|---|---|---|
| *Bulk* | | | |
| Theoretical: | | | |
| PBEsol | 3.89 | 3.66 | 14.84 |
| PBEsol + vdW[45] | 3.83 | 3.67 | 13.25 |
| Experimental: | 3.96 | 3.72 | 13.86 |
| *Monolayer* | | | |
| Theoretical | | | |
| PBEsol | 3.88 | 3.65 | - |
| PBEsol + vdW[45] | 3.85 | 3.64 | |

**Table 1.** MoO$_3$ lattice parameters for both bulk and monolayer. Theoretical results are compared with previously reported experimental values.[27, 28] Note that van der Waals correction has a sizable influence in the *c* parameter in the bulk structure, as has previously shown in Ref.[46]. In comparison, in monolayer calculations the effect of Van der Waals correction is negligible, due to the stronger bond within the layers.

AUTHOR INFORMATION

**Corresponding Authors**

*aday.molina@uam.es

*andres.castellanos@imdea.org

ACKNOWLEDGMENT

A.C-G. acknowledges financial support from the BBVA Foundation through the fellowship "I Convocatoria de Ayudas Fundacion BBVA a Investigadores, Innovadores y Creadores Culturales", from the MINECO (Ramón y Cajal 2014 program, RYC-2014-01406) and from the MICINN (MAT2014-58399-JIN). A.J.M-M., G.R-B. and N.A. acknowledge the support of the MICCINN/MINECO (Spain) through the programs MAT2014-57915-R, BES-2012-057346 and FIS2011-23488, Comunidad de Madrid (Spain) through the programs NANOBIOMAGNET (s2009/MAT-1726) and S2013/MIT-3007 (MAD2D). J.I. and H.S.J.vdZ. acknowledgethe Dutch organization for Fundamental Research on Matter (FOM) and by the Ministry of Education, Culture, and Science (OCW). J.L. and J.F-R. acknowledge Marie Curie ITN SPINOGRAH (607904-13). FYB acknowledges financial support from the Swiss National Science Foundation (Ambizione grant PZ00P2_161327). E.M.P. acknowledges financial support from the European Research Council (StG-307609-MINT) and the MINECO of Spain (CTQ2014-60541-P).

REFERENCES

1.      Novoselov, K. S.; Geim, A. K.; Morozov, S. V.; Jiang, D.; Zhang, Y.; Dubonos, S. V.; Grigorieva, I. V.; Firsov, A. A., Electric Field Effect in Atomically Thin Carbon Films. *Science* **2004,** 306, 666-669.

2. Radisavljevic, B.; Radenovic, A.; Brivio, J.; Giacometti, V.; Kis, A., Single-layer MoS2 transistors. *Nat. Nanotechnol.* **2011,** 6, 147-150.
3. Reina, A.; Jia, X.; Ho, J.; Nezich, D.; Son, H.; Bulovic, V.; Dresselhaus, M. S.; Kong, J., Large Area, Few-Layer Graphene Films on Arbitrary Substrates by Chemical Vapor Deposition. *Nano Lett.* **2008,** 9, 30-35.
4. Zhan, Y.; Liu, Z.; Najmaei, S.; Ajayan, P. M.; Lou, J., Large-Area Vapor-Phase Growth and Characterization of MoS2 Atomic Layers on a SiO2 Substrate. *Small* **2012,** 8, 966-971.
5. Kalantar-zadeh, K.; Tang, J.; Wang, M.; Wang, K. L.; Shailos, A.; Galatsis, K.; Kojima, R.; Strong, V.; Lech, A.; Wlodarski, W.; Kaner, R. B., Synthesis of nanometre-thick MoO3 sheets. *Nanoscale* **2010,** 2, 429-433.
6. Bouzidi, A.; Benramdane, N.; Tabet-Derraz, H.; Mathieu, C.; Khelifa, B.; Desfeux, R., Effect of substrate temperature on the structural and optical properties of MoO3 thin films prepared by spray pyrolysis technique. *Mater. Sci. Eng., B* **2003,** 97, 5-8.
7. Balendhran, S.; Deng, J.; Ou, J. Z.; Walia, S.; Scott, J.; Tang, J.; Wang, K. L.; Field, M. R.; Russo, S.; Zhuiykov, S.; Strano, M. S.; Medhekar, N.; Sriram, S.; Bhaskaran, M.; Kalantar-zadeh, K., Enhanced Charge Carrier Mobility in Two-Dimensional High Dielectric Molybdenum Oxide. *Adv. Mater.* **2013,** 25, 109-114.
8. Balendhran, S.; Walia, S.; Nili, H.; Ou, J. Z.; Zhuiykov, S.; Kaner, R. B.; Sriram, S.; Bhaskaran, M.; Kalantar-zadeh, K., Two-Dimensional Molybdenum Trioxide and Dichalcogenides. *Adv. Funct. Mater.* **2013,** 23, 3952-3970.
9. Xiang, D.; Han, C.; Zhang, J.; Chen, W., Gap States Assisted MoO3 Nanobelt Photodetector with Wide Spectrum Response. *Sci. Rep.* **2014,** 4, 4891.
10. Yu, T.; Zhu, Y. W.; Xu, X. J.; Shen, Z. X.; Chen, P.; Lim, C. T.; Thong, J. T. L.; Sow, C. H., Controlled Growth and Field-Emission Properties of Cobalt Oxide Nanowalls. *Adv. Mater.* **2005,** 17, 1595-1599.
11. Yan, B.; Zheng, Z.; Zhang, J.; Gong, H.; Shen, Z.; Huang, W.; Yu, T., Orientation Controllable Growth of MoO3 Nanoflakes: Micro-Raman, Field Emission, and Birefringence Properties. *J. Phys. Chem. C* **2009,** 113, 20259-20263.
12. Koma, A.; Sunouchi, K.; Miyajima, T., Fabrication and characterization of heterostructures with subnanometer thickness. *Microelectron. Eng.* **1984,** 2, 129-136.
13. Wang, Q.; Safdar, M.; Xu, K.; Mirza, M.; Wang, Z.; He, J., Van der Waals Epitaxy and Photoresponse of Hexagonal Tellurium Nanoplates on Flexible Mica Sheets. *ACS Nano* **2014,** 8, 7497-7505.
14. Li, H.; Cao, J.; Zheng, W.; Chen, Y.; Wu, D.; Dang, W.; Wang, K.; Peng, H.; Liu, Z., Controlled Synthesis of Topological Insulator Nanoplate Arrays on Mica. *JACS* **2012,** 134, 6132-6135.
15. Utama, M. I. B.; Belarre, F. J.; Magen, C.; Peng, B.; Arbiol, J.; Xiong, Q., Incommensurate van der Waals Epitaxy of Nanowire Arrays: A Case Study with ZnO on Muscovite Mica Substrates. *Nano Lett.* **2012,** 12, 2146-2152.
16. Utama, M. I. B.; Zhang, Q.; Jia, S.; Li, D.; Wang, J.; Xiong, Q., Epitaxial II–VI Tripod Nanocrystals: A Generalization of van der Waals Epitaxy for Nonplanar Polytypic Nanoarchitectures. *ACS Nano* **2012,** 6, 2281-2288.
17. Wang, Q.; Xu, K.; Wang, Z.; Wang, F.; Huang, Y.; Safdar, M.; Zhan, X.; Wang, F.; Cheng, Z.; He, J., van der Waals Epitaxial Ultrathin Two-Dimensional Nonlayered Semiconductor for Highly Efficient Flexible Optoelectronic Devices. *Nano Lett.* **2015,** 15, 1183-1189.

18. Wang, Z.; Safdar, M.; Mirza, M.; Xu, K.; Wang, Q.; Huang, Y.; Wang, F.; Zhan, X.; He, J., High-performance flexible photodetectors based on GaTe nanosheets. *Nanoscale* **2015,** 7, 7252-7258.
19. Castellanos-Gomez, A.; Buscema, M.; Molenaar, R.; Singh, V.; Janssen, L.; van der Zant, H. S. J.; Steele, G. A., Deterministic transfer of two-dimensional materials by all-dry viscoelastic stamping. *2D Mater.* **2014,** 1, 011002.
20. Blake, P.; Hill, E. W.; Castro Neto, A. H.; Novoselov, K. S.; Jiang, D.; Yang, R.; Booth, T. J.; Geim, A. K., Making graphene visible. *Appl. Phys. Lett.* **2007,** 91, 063124.
21. Castellanos-Gomez, A.; Agraït, N.; Rubio-Bollinger, G., Optical identification of atomically thin dichalcogenide crystals. *Appl. Phys. Lett.* **2010,** 96, 213116.
22. Baltrusaitis, J.; Mendoza-Sanchez, B.; Fernandez, V.; Veenstra, R.; Dukstiene, N.; Roberts, A.; Fairley, N., Generalized molybdenum oxide surface chemical state XPS determination via informed amorphous sample model. *Appl. Surf. Sci.* **2015,** 326, 151-161.
23. Simchi, H.; McCandless, B. E.; Meng, T.; Boyle, J. H.; Shafarman, W. N., Characterization of reactively sputtered molybdenum oxide films for solar cell application. *J. Appl. Phys.* **2013,** 114, 013503.
24. Mestl, G.; Srinivasan, T. K. K., Raman Spectroscopy of Monolayer-Type Catalysts: Supported Molybdenum Oxides. *Cataly. Rev.* **1998,** 40, 451-570.
25. Taka-aki, Y.; Keisuke, Y.; Yuhei, H.; Tomohiro, H.; Fumio, O.; Masahiko, H., Probing edge-activated resonant Raman scattering from mechanically exfoliated 2D MoO 3 nanolayers. *2D Mater.* **2015,** 2, 035004.
26. Spevack, P. A.; McIntyre, N. S., Thermal reduction of molybdenum trioxide. *J. Phys. Chem.* **1992,** 96, 9029-9035.
27. Sitepu, H., Texture and structural refinement using neutron diffraction data from molybdite (MoO 3) and calcite (CaCO 3) powders and a Ni-rich Ni 50.7 Ti 49.30 alloy. *Powder Diffr.* **2009,** 24, 315-326.
28. Negishi, H.; Negishi, S.; Kuroiwa, Y.; Sato, N.; Aoyagi, S., Anisotropic thermal expansion of layered MoO₃ crystals. *Phys. Rev. B* **2004,** 69, 064111.
29. Tran, F.; Blaha, P., Accurate Band Gaps of Semiconductors and Insulators with a Semilocal Exchange-Correlation Potential. *Phys. Rev. Lett.* **2009,** 102, 226401.
30. Chen, Z.; Santoso, I.; Wang, R.; Xie, L. F.; Mao, H. Y.; Huang, H.; Wang, Y. Z.; Gao, X. Y.; Chen, Z. K.; Ma, D.; Wee, A. T. S.; Chen, W., Surface transfer hole doping of epitaxial graphene using MoO3 thin film. *Appl. Phys. Lett.* **2010,** 96, 213104.
31. Russell, S. A. O.; Cao, L.; Qi, D.; Tallaire, A.; Crawford, K. G.; Wee, A. T. S.; Moran, D. A. J., Surface transfer doping of diamond by MoO3: A combined spectroscopic and Hall measurement study. *Appl. Phys. Lett.* **2013,** 103, 202112.
32. Zheng, Q.; Huang, J.; Cao, S.; Gao, H., A flexible ultraviolet photodetector based on single crystalline MoO3 nanosheets. *J. Mater. Chem. C* **2015,** 3, 7469-7475.
33. Hu, P.; Wang, L.; Yoon, M.; Zhang, J.; Feng, W.; Wang, X.; Wen, Z.; Idrobo, J. C.; Miyamoto, Y.; Geohegan, D. B.; Xiao, K., Highly Responsive Ultrathin GaS Nanosheet Photodetectors on Rigid and Flexible Substrates. *Nano Lett.* **2013,** 13, 1649-1654.
34. Xu, K.; Wang, Z.; Wang, F.; Huang, Y.; Wang, F.; Yin, L.; Jiang, C.; He, J., Ultrasensitive Phototransistors Based on Few-Layered HfS2. *Adv. Mater.* **2015,** 27, 7881-7887.
35. Nakayama, Y.; Morii, K.; Suzuki, Y.; Machida, H.; Kera, S.; Ueno, N.; Kitagawa, H.; Noguchi, Y.; Ishii, H., Origins of Improved Hole-Injection Efficiency by the Deposition of MoO3 on the Polymeric Semiconductor Poly(dioctylfluorene-alt-benzothiadiazole). *Adv. Funct. Mater.* **2009,** 19, 3746-3752.

36. Kröger, M.; Hamwi, S.; Meyer, J.; Riedl, T.; Kowalsky, W.; Kahn, A., Role of the deep-lying electronic states of MoO3 in the enhancement of hole-injection in organic thin films. *Appl. Phys. Lett.* **2009,** 95, 123301.

37. Kröger, M.; Hamwi, S.; Meyer, J.; Riedl, T.; Kowalsky, W.; Kahn, A., P-type doping of organic wide band gap materials by transition metal oxides: A case-study on Molybdenum trioxide. *Org. Electron.* **2009,** 10, 932-938.

38. Meyer, J.; Kidambi, P. R.; Bayer, B. C.; Weijtens, C.; Kuhn, A.; Centeno, A.; Pesquera, A.; Zurutuza, A.; Robertson, J.; Hofmann, S., Metal Oxide Induced Charge Transfer Doping and Band Alignment of Graphene Electrodes for Efficient Organic Light Emitting Diodes. *Sci. Rep.* **2014,** 4, 5380.

39. Lin, J.; Zhong, J.; Zhong, S.; Li, H.; Zhang, H.; Chen, W., Modulating electronic transport properties of MoS2 field effect transistor by surface overlayers. *Appl. Phys. Lett.* **2013,** 103, 063109.

40. Fontana, M.; Deppe, T.; Boyd, A. K.; Rinzan, M.; Liu, A. Y.; Paranjape, M.; Barbara, P., Electron-hole transport and photovoltaic effect in gated MoS2 Schottky junctions. *Sci. Rep.* **2013,** 3, 1634.

41. Aballe, L.; Foerster, M.; Pellegrin, E.; Nicolas, J.; Ferrer, S., The ALBA spectroscopic LEEM-PEEM experimental station: layout and performance. *J. Synchrotron Radia.* **2015,** 22, 745-752.

42. Giannozzi, P.; Baroni, S.; Bonini, N.; Calandra, M.; Car, R.; Cavazzoni, C.; Ceresoli, D.; Chiarotti, G. L.; Cococcioni, M.; Dabo, I.; Corso, A. D.; Gironcoli, S. d.; Fabris, S.; Fratesi, G.; Gebauer, R.; Gerstmann, U.; Gougoussis, C.; Kokalj, A.; Lazzeri, M.; Martin-Samos, L.; Marzari, N.; Mauri, F.; Mazzarello, R.; Paolini, S.; Pasquarello, A.; Paulatto, L.; Sbraccia, C.; Scandolo, S.; Sclauzero, G.; Seitsonen, A. P.; Smogunov, A.; Umari, P.; Wentzcovitch, R. M., QUANTUM ESPRESSO: a modular and open-source software project for quantum simulations of materials. *J. Phys.: Condens. Matter* **2009,** 21, 395502.

43. Elk, http://elk.sourceforge.net/.

44. Perdew, J. P.; Ruzsinszky, A.; Csonka, G. I.; Vydrov, O. A.; Scuseria, G. E.; Constantin, L. A.; Zhou, X.; Burke, K., Restoring the Density-Gradient Expansion for Exchange in Solids and Surfaces. *Phys. Rev. Lett.* **2008,** 100, 136406.

45. Grimme, S., Semiempirical GGA-type density functional constructed with a long-range dispersion correction. *J. Comput. Chem.* **2006,** 27, 1787-1799.

46. Ding, H.; Ray, K. G.; Ozolins, V.; Asta, M., Structural and vibrational properties of alpha-MoO3 from van der Waals corrected density functional theory calculations. *Phys. Rev. B* **2012,** 85, 012104.

**Supporting Information.**

**Centimeter-scale synthesis of ultrathin layered MoO$_3$ by van der Waals epitaxy**

*Aday J. Molina-Mendoza, Jose Luis Lado, Joshua O. Island, Miguel Angel Niño, Lucía Aballe, Michael Foerster, Flavio Y. Bruno, Alejandro López-Moreno, Luis Vaquero-Garzon, Herre S. J. van der Zant, Gabino Rubio-Bollinger, Nicolas Agraït, Emilio M. Pérez, Joaquin Fernandez-Rossier, and Andres Castellanos-Gomez*

In this supporting information we include the following content:

- MoO$_3$ grown on different substrates
- MoO$_3$ transferred to different substrates
- Low energy electron microscopy characterization
- Optical conductivity calculations of MoO$_3$
- Large-scale devices

**MoO3 grown on different substrates**

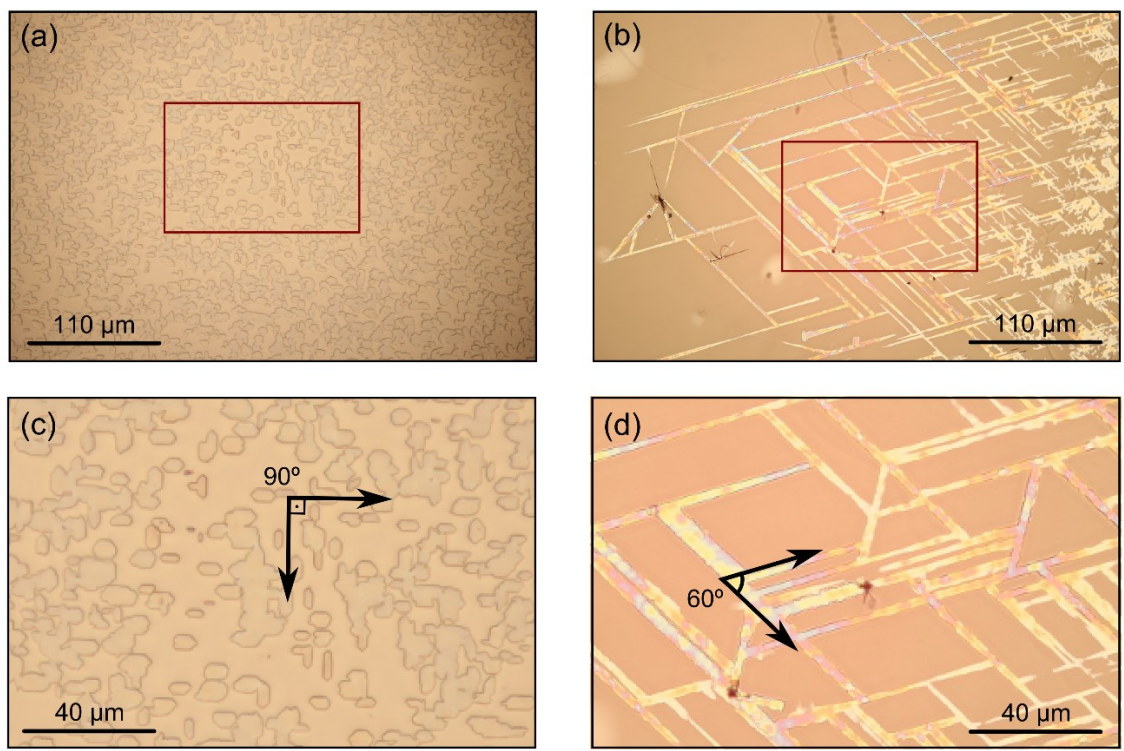

**Figure S1** Optical microscopy images of MoO3 grown on **(a)** SrTiO2 and **(b)** mica using the method illustrated in Figure 1. **(c)** and **(d)** show optical microscopy images of MoO3 grown on SrTiO2 and mica, respectively, corresponding to the areas highlighted by red squares in (a) and (b). As indicated by the arrows, the growth directions of the crystals depend on the crystallographic structure of the target substrate: 90º for SrTiO2 and 60º for mica.

SrTiO3 substrates

We have used (001)-oriented SrTiO3 (STO) substrates from Crystec GmbH with a miscut angle lower than 0.5°. STO presents a cubic perovskite structure at room temperature with a lattice parameter a = 0.3905 nm. The polished surface of (001) substrates presents a mixture of both possible terminations SrO and $TiO_2$ planes and rough step edges. In order to obtain the $TiO_2$ surface terminated substrate with straight step and terrace structure shown in Figure S2 we followed a standard etching-annealing procedure.[1] The substrates were ultrasonically soaked in demineralized water for 10 min followed by a dip during 40 sec in a buffered HF solution ($NH_4F$:HF = 87.5:12.5, obtained from Merck) and rinsed once again in demineralized water. A final annealing step during 2 hours under oxygen flow at 950°C ensures the recrystallization of the surface.

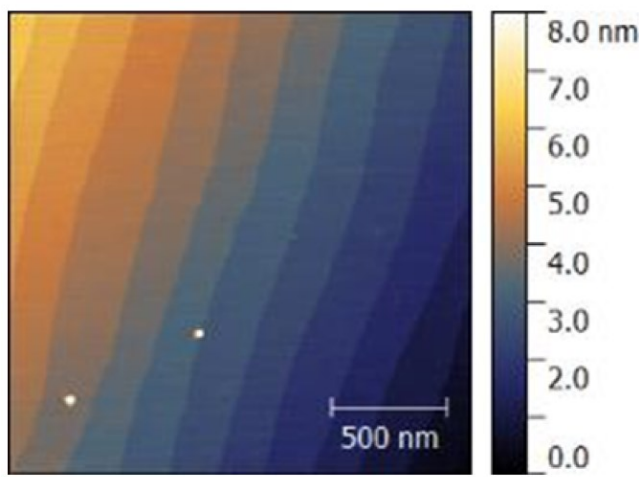

**Figure S2** AFM topographic image of a $TiO_2$ terminated $SrTiO_3$ substrate surface.

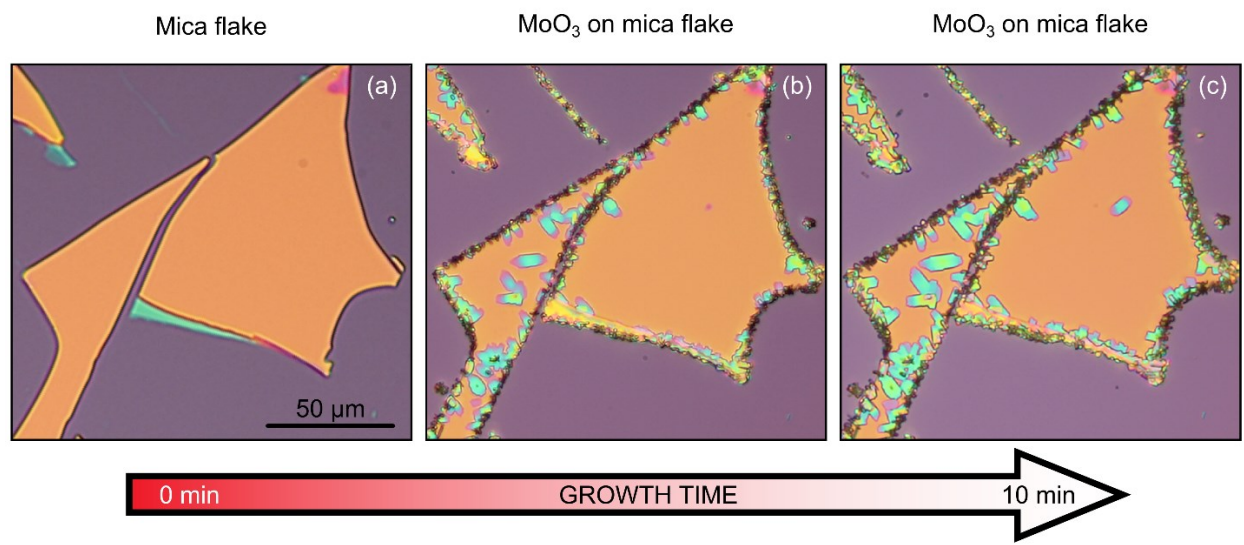

**Figure S3** MoO$_3$ grown on an exfoliated mica flake deposited on a SiO$_2$ substrate for different growth times. **(a)** Mica flake deposited on a SiO$_2$ substrate. **(b)** MoO$_3$ grown on the mica flake after 5 min of growth. **(c)** MoO3 grown on the mica flake after 10 min of growth.

**MoO$_3$ transferred to different substrates**

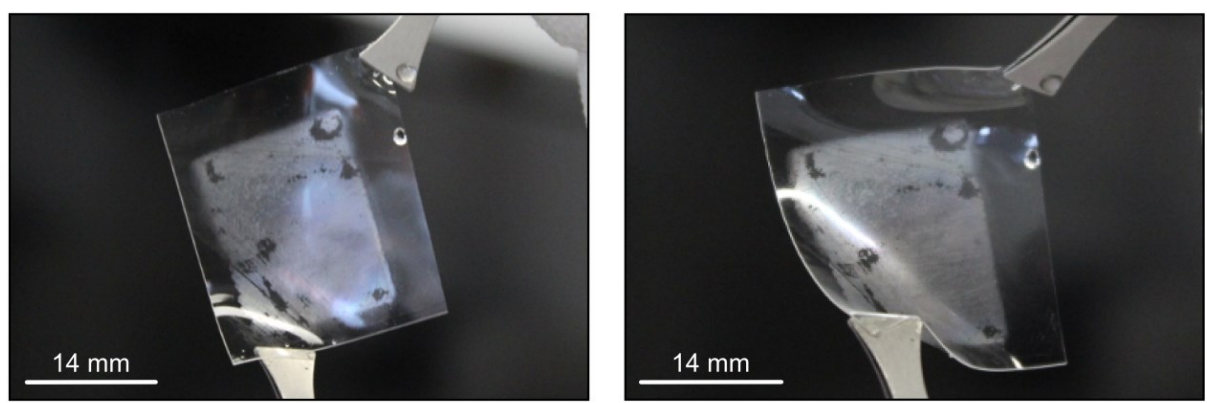

**Figure S4** Photographs of MoO$_3$ extended layer (grown thick on purpose for the pictures) on a PDMS stamp after being detached from the mica substrate by the method illustrated in Figure 4 of the main text. As shown in the pictures, MoO3 can be stretched together with the substrate, and therefore is a suitable material for flexible and transparent electronics.

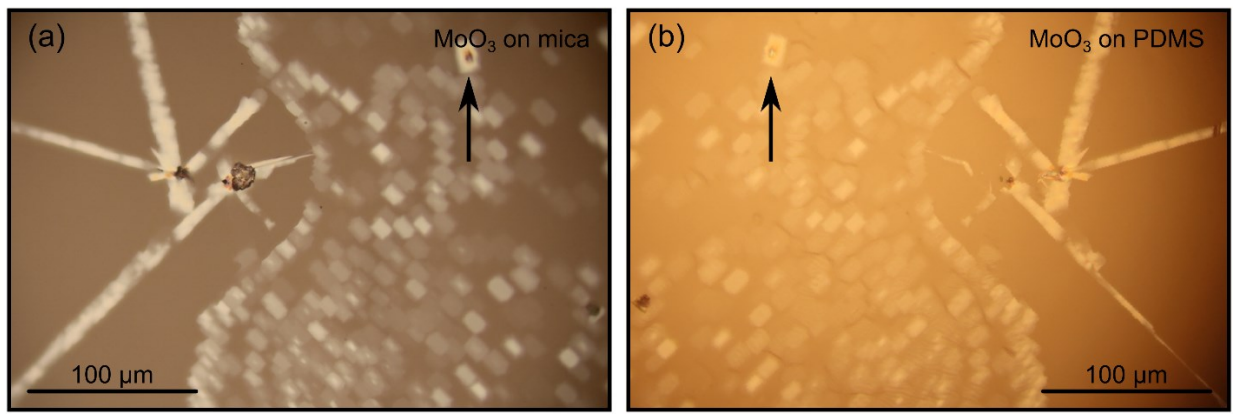

**Figure S5 (a)** Optical microscopy image of the as-grown MoO3 crystals on a mica substrate. **(b)** The same MoO3 crystals as in (a) after been detached from the mica substrate with a PDMS stamp.

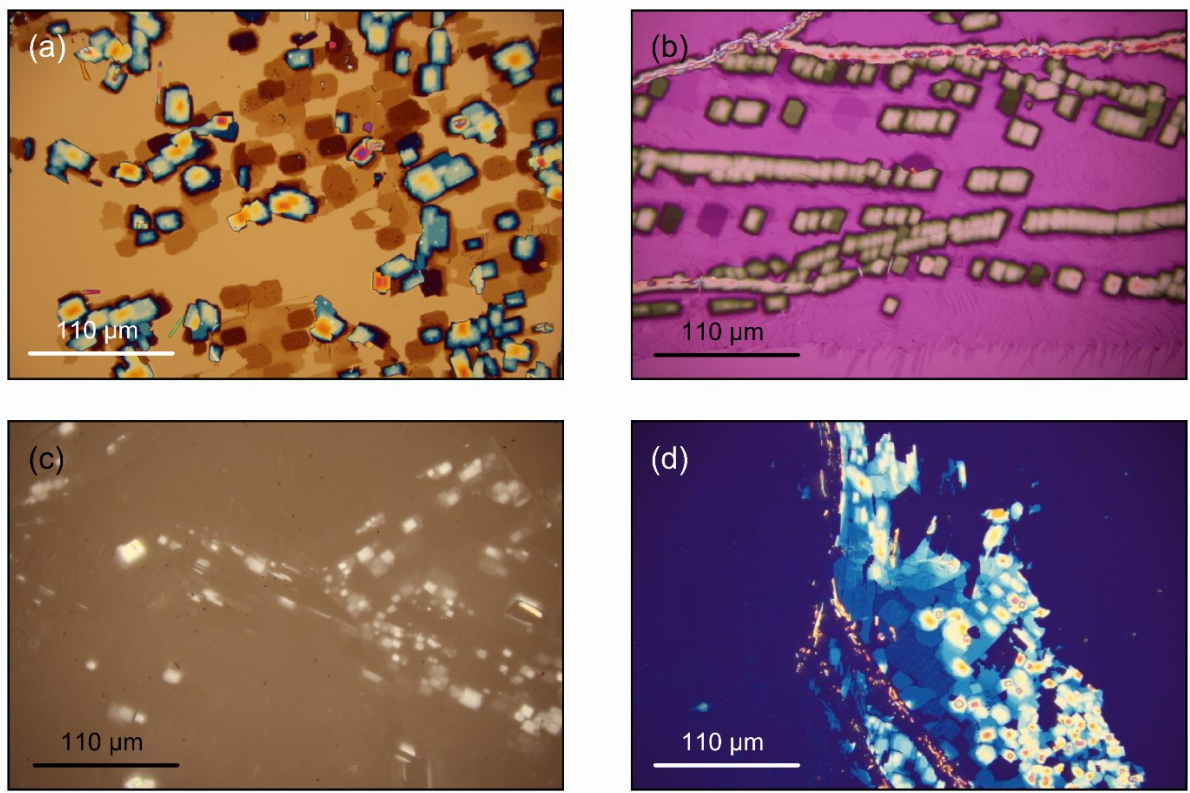

**Figure S6** Optical microscopy images of MoO$_3$ crystals transferred (after growth on mica) to **(a)** an evaporated thin film of Pt on SiO$_2$, **(b)** an evaporated thin film of Au on SiO$_2$, **(c)** a quartz substrate and **(d)** a Si$_3$N$_4$ substrate.

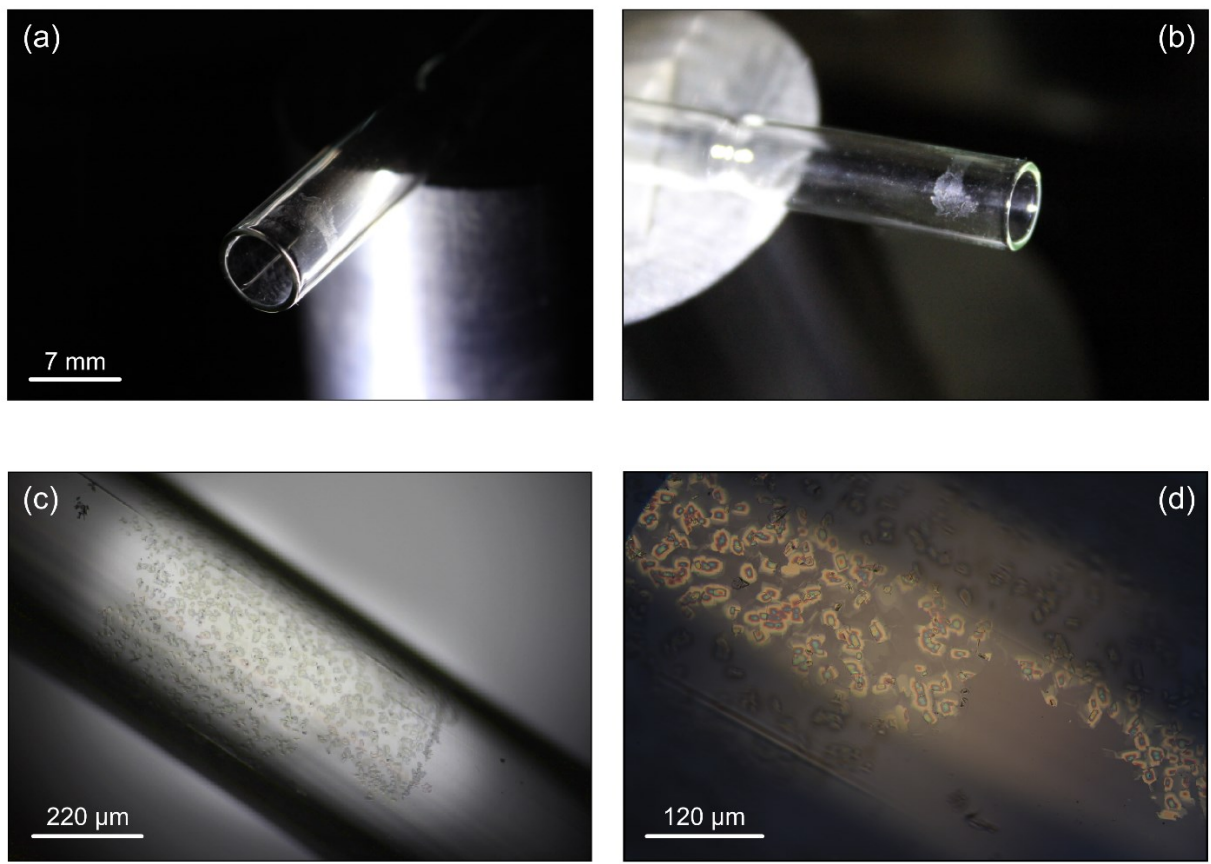

**Figure S7 (a)** and **(b)** Photographs of MoO$_3$ (grown thick on purpose for the pictures) transferred to a rounded and transparent glass substrate. **(c)** and **(d)** show optical microscopy images of the photographs in (a) and (b).

**Low energy electron microscopy characterization of MoO₃ extended layers**

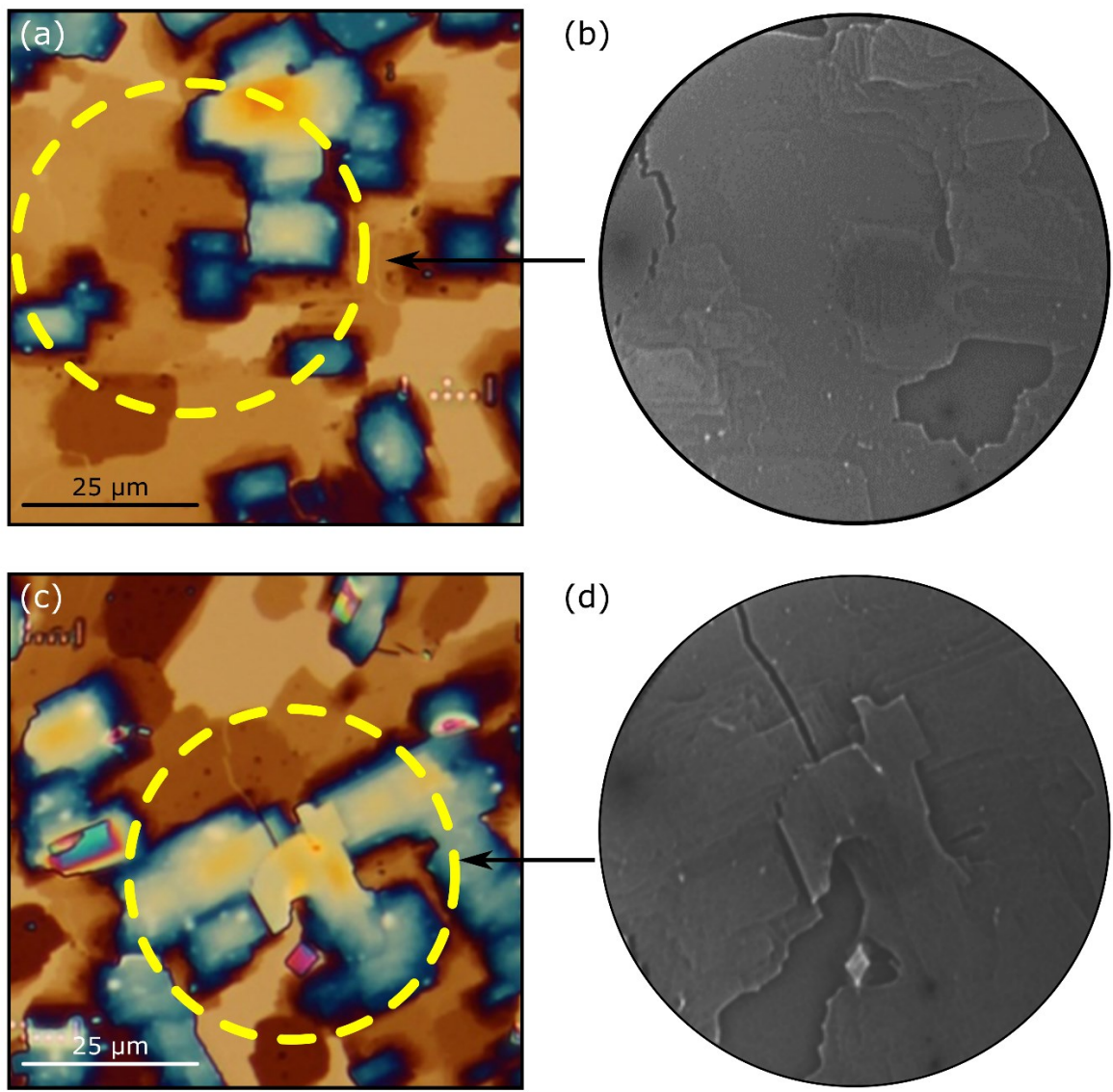

**Figure S8 (a)** Optical microscopy image of MoO₃ transferred on a Au substrate and **(b)** LEEM image corresponding to the area highlighted by a dashed yellow circle. **(c)** and **(d)** Same as in (a) and (b) for another area in the same sample.

**Optical conductivity calculations of MoO₃**

In order to compute optical absorption, we compute the optical conductivity of MoO₃ using the Random Phase Approximation (RPA) in the q→0 limit, starting from the TB-mBJ potential calculation. The resulting optical spectra (plotted in Figure S9a) shows the same trend as the band structure calculations, i.e., there is no much difference in the optical absorption between monolayer and bulk at this level of approximation. The analysis of the density of states (Figure S9b) shows that the lowest optical transitions involve excitations from the top of the valence band, with a dominant weight in oxygen *p* orbitals, to the bottom of the conduction band which formed mainly by Mo *d* orbitals.

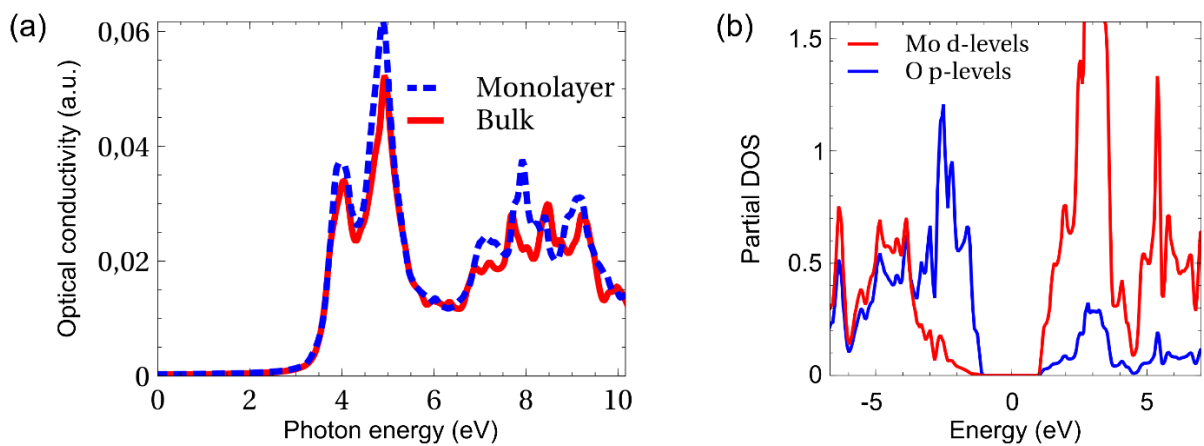

**Figure S9 (a)** Calculated optical absorption for both monolayer and bulk MoO₃. **(b)** Calculated optical conductivity for both monolayer and bulk MoO₃ in agreement with Ref.2

**Large-scale devices**

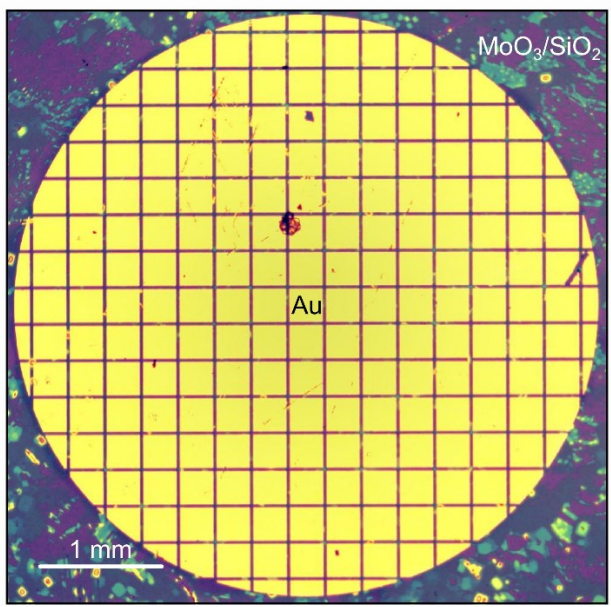

**Figure S10** Optical microscopy images of the fabricated $MoO_3$-based devices. The Ti/Au electrodes (5 nm/ 70 nm thick) are evaporated via shadow mask on the extended layer of $MoO_3$ transferred to a $SiO_2$ substrate.

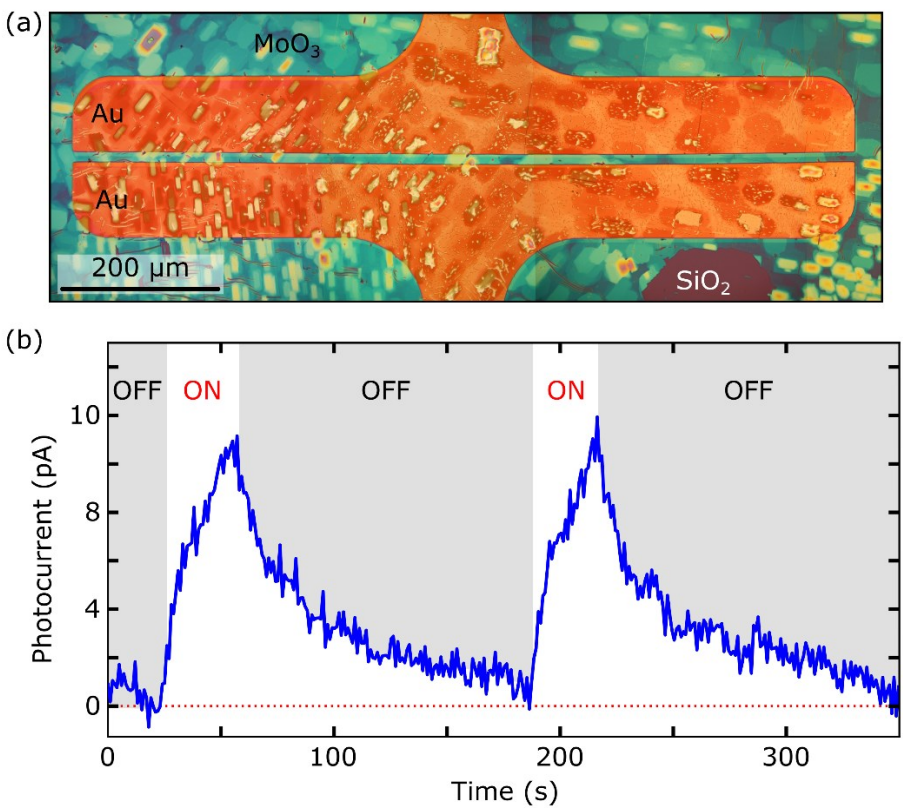

**Figure S11 (a)** Large-scale $MoO_3$ photodetector fabricated by deterministic transfer of $MoO_3$ on pre-patterned Au electrodes on a $SiO_2$/Si substrate. **(b)** Photocurrent time response of the device shown in (a). The shaded areas indicate the period where the light was switched off and the white areas the period while the light was on. From this curve we obtain a rising time of ~ 20 s and a fall time of ~ 68 s.

Supporting Information references
1. Koster, G.; Kropman, B. L.; Rijnders, G. J. H. M.; Blank, D. H. A.; Rogalla, H., Quasi-ideal strontium titanate crystal surfaces through formation of strontium hydroxide. *Appl. Phys. Lett.* **1998,** 73, 2920-2922.
2. Scanlon, D. O.; Watson, G. W.; Payne, D. J.; Atkinson, G. R.; Egdell, R. G.; Law, D. S. L., Theoretical and Experimental Study of the Electronic Structures of MoO3 and MoO2. *J. Phys. Chem. C* **2010,** 114, 4636-4645.